\documentclass[aps,pra,reprint,superscriptaddress,footinbib,longbibliography]{revtex4-2}

\usepackage{amssymb}
\usepackage{amsmath}
\usepackage{amsfonts}
\usepackage{xcolor}
\usepackage{graphicx}
\usepackage{bm}
\usepackage{physics}
\usepackage{stackengine}
\usepackage{enumitem}
\usepackage{mathtools}
\usepackage[normalem]{ulem}
\usepackage{verbatim}

\usepackage[unicode]{hyperref}
\hypersetup{
   unicode=true,          
   plainpages=false,
   colorlinks=true,       
   citecolor=blue,        
}

\allowdisplaybreaks

\newcommand{\sgn}{\mathrm{sgn}}

\begin{document}

\title{Odd-frequency pair correlations in the trapped
fermionic Tonks-Girardeau gas}

\author{M. Wong}
\affiliation{School of Chemical and Physical Sciences,
Victoria University of Wellington, PO Box 600, Wellington
6140, New Zealand}

\author{K. Thompson}
\email{kadin.thompson@vuw.ac.nz}
\affiliation{School of Chemical and Physical Sciences,
Victoria University of Wellington, PO Box 600, Wellington
6140, New Zealand}

\author{M. Governale}
\affiliation{School of Chemical and Physical Sciences,
Victoria University of Wellington, PO Box 600, Wellington
6140, New Zealand}

\author{J. Brand}
\affiliation{Centre for Theoretical Chemistry and
Physics, New Zealand Institute for Advanced Study, Massey
University,Private Bag 102904, North Shore, Auckland
0745, New Zealand}
\affiliation{The Dodd-Walls Centre for Photonic and
Quantum Technologies, New Zealand}

\author{U. Z\"ulicke}
\email{uli.zuelicke@vuw.ac.nz}
\affiliation{School of Chemical and Physical Sciences,
Victoria University of Wellington, PO Box 600, Wellington
6140, New Zealand}
\affiliation{The Dodd-Walls Centre for Photonic and
Quantum Technologies, New Zealand}

\date{\today}

\begin{abstract}
The fermionic Tonks-Girardeau gas is an exactly solvable
model realizing a conventional macroscopic condensate of
$p$-wave Cooper pairs. Here we demonstrate that, when
confined by a trapping potential, it also hosts
unconventional odd-frequency pairing, i.e., pair
correlations with $s$-wave symmetry that are only present
with a finite time delay $t$ between fermions forming a
pair. Quantitative results for fixed particle number are
obtained using a generalization of Yang's theory of 
off-diagonal long-range order to define a $t$-dependent
order parameter that is antisymmetric under combined $t$
inversion and exchange of fermion-pair indices.
Odd-frequency pairing in the fermionic Tonks-Girardeau
gas is found to be a mesoscopic effect, localized at the
system's boundary and therefore nonextensive
thermodynamically. Our results show that this elusive
hidden order is accessible in systems having fixed
particle number, suggesting new avenues towards its
experimental realization and further detailed study.
\end{abstract}

\maketitle

\section{Introduction}
\label{sec:intro}

Odd-frequency pairing~\cite{Berezinskii1974,Linder2019}
is an intriguing, but at the same time quite elusive,
type of hidden order~\cite{Aeppli2020}. While various
mechanisms for its realization have been proposed
theoretically~\cite{Balatsky1992,Coleman1994,Coleman1995,
Abrahams1995,Bergeret2005,Eschrig2007,Tanaka2007,
Matsumoto2012,Black-Schaffer2013,Sothmann2014}, only
indirect experimental signatures have been
observed~\cite{DiBernardo2015,Pal2017,Krieger2020,
Perrin2020}, and the possible existence of such an
unconventional pair condensate has been challenged on 
fundamental grounds~\cite{Heid1995,Solenov2009,
Kusunose2011,Fominov2015,Schrodi2021,Pimenov2022,
Langmann2022}. Studies of exactly solvable model systems
could provide conclusive answers to the numerous open
questions concerning the character of odd-frequency
pairing and its potential realizability in nature, but
very few of these are currently
available~\cite{deFarias2020,Coleman2022}. Our present
work opens up a promising new avenue by revealing the
existence of mesoscopic odd-frequency pair correlations
in the fermionic Tonks-Girardeau (FTG) gas, an exactly
solvable model~\cite{Girardeau2004} that has been
extensively studied~\cite{Girardeau2005,Bender2005,
Hao2007,Koscik2023,Sabater2024,Girardeau2006,
Minguzzi2006,Sabater2025}. In the process, we demonstrate
the practical utility of a recently developed
particle-number-conserving formalism for describing
odd-frequency pairing~\cite{Thompson2024}. We proceed by
introducing the theoretical approach underpinning this
work, before presenting our results in the subsequent
sections.

The concept of off-diagonal long-range order (ODLRO) was
developed~\cite{Penrose1956,Yang1962} to describe the
emergence of macroscopic quantum condensates, which
exhibit the phenomena of superfluidity and
superconductivity~\cite{Leggett2006}. It centers around
the properties of reduced density
matrices~\cite{Coleman1963}. Thus it is applicable to
systems with fixed particle number~\cite{Yukalov2025}, in
contrast to the popular concept of spontaneous symmetry
breaking of the $U(1)$ gauge symmetry that underlies the
typical mean-field approaches~\cite{Anderson1966,
Bruus2004} and requires an explicit breaking of
particle-number conservation. The fact that ODLRO theory
enables particle-number conservation makes it ideally
suited to reveal macroscopic order in numerical
studies~\cite{vdLinden1992,Ebling2021a,Pakrouski2026,
Cufar2026,Brand2026}, as well as for gaining a deeper
understanding of instructive particle-number-conserving
model systems~\cite{Yang1989,Girardeau2006,Minguzzi2006,
Kraus2009,Knight2022,Koscik2023,Sabater2024}.

Recently, the ODLRO formalism was extended to discuss
odd-frequency pairing within a particle-number-conserving
framework~\cite{Thompson2024}. This approach is based on
consideration of the $t$-dependent two-body correlation
matrix (T2bCM) that, for the specific context of our
present work, has the form
\begin{align}\label{eq:T2bCMgen}
& \rho_2(x,y; x',y';t) = \nonumber \\[0.1cm]
& \hspace{0.1cm} \Big\langle\Psi_0\Big| c^\dag_x\,\exp
(-i\,\frac{t}{\hbar}\, H)\, c^\dag_y\, c_{y'}\, \exp(i\,
\frac{t}{\hbar}\, H)\, c_{x'}\Big|\Psi_0\Big\rangle\,\, .
\end{align}
Here $c^\dag_x$ and $c_x$ indicate the second-quantized
operators for creating and annihilating a fermion at
position $x$, $\ket{\Psi_0}$ is the system's $N$-particle
ground state, $t$ measures the time delay between
fermions forming a Cooper pair, and $H$ denotes the
particle-number-conserving many-body Hamiltonian.

The T2bCM generalizes the familiar two-body reduced
density matrix~\footnote{Appendix~\ref{app:2bRDMconv}
discusses in more detail the usual conventions for
defining the two-body reduced density matrix.}
\begin{align}\label{eq:2bRDMgen}
\rho_2(x,y; x',y') &= \bra{\Psi_0} c^\dag_x\, c^\dag_y\,
c_{y'}\, c_{x'} \ket{\Psi_0} \equiv \rho_2(x,y; x',y';0)
\end{align}
that underpins the ODLRO description of even-frequency
pairing~\cite{Yang1962}. In particular, it has been
shown~\cite{Thompson2024} that the T2bCM is Hermitian and
positive-semidefinite, and that it has a $t$-independent
trace $\mathrm{Tr}\{\rho_2(x,y; x', y';t)\}=N(N-1)$.
As a result, the T2bCM's spectral decomposition
\begin{align}\label{eq:T2bCMdiag}
\rho_2(x, y; x', y'; t) = \sum_\alpha n_\alpha(t)\big[
\chi_\alpha(x, y; t)\big]^* \chi_\alpha(x', y'; t)
\end{align}
provides the basis for identifying macroscopic order for
$t>0$ based on the conventional Yang
criterion~\cite{Yang1962}: that there exists an
eigenvalue $n_0(t)$ of the T2bCM that scales with the
particle number $N$. The function 
\begin{align}\label{eq:OPdef}
\phi_0(x, y; t) &= \sqrt{n_0(t)}\,\,
\chi_0(x, y; t)
\end{align}
is then the order parameter representing the fermion-pair
condensate that gives rise to superfluidity. Here, the
normalized eigenvector $\chi_0(x, y; t)$ can be
interpreted as the macroscopically occupied Cooper-pair
state.

It is mandated by Fermi statistics that any order
parameter associated with nonvanishing $n_0(0)$ is
antisymmetric under index permutation $P$, i.e.,
$P\phi_0(x, y; 0) = \phi_0(y, x; 0) = -\phi_0(x, y; 0)$.
This condition is relaxed for $t\ne 0$ so that
$\phi_0(x, y; t)$ generally has both symmetric and
antisymmetric parts~\cite{Thompson2024},
\begin{subequations}
\begin{align}\label{eq:symmPhi}
S\phi_0(x, y; t) &= \frac{1}{2} \left[ \phi_0 (x, y; t)
+ \phi_0(y, x; t)\right]\, , \\[0.1cm] \label{eq:asPhi}
A\phi_0(x, y; t) &= \frac{1}{2} \left[ \phi_0(x, y; t)
- \phi_0(y, x; t)\right]\, .
\end{align}
\end{subequations}
As $S\phi_0(x, y; t)$ vanishes for $t\to 0$, it
embodies pair correlations that grow with the time-delay
parameter $t$, which is the distinctive feature of
odd-in-time, also called odd-frequency,
pairing~\cite{Berezinskii1974,Linder2019}.

More rigorous insights about how odd-frequency pairing
emerges from the interplay of symmetry under $P$ and
under time ($t$) reversal $T$~\cite{Linder2019} can be
gained by considering the time-ordered two-body
correlation matrix
\begin{widetext}
\begin{align}\label{eq:TO2bCM}
& \rho_2^{\mathrm{T}}(x,y; x',y';t) = \theta(t)\, \rho_2
(x, y; x',y';t) + \theta(-t)\, \rho_2(y, x; y', x'; -t)
\,\, , 
\end{align}
where $\theta(t)$ is the Heaviside step function
satisfying $\theta(0)=1/2$. The time-ordered two-body
correlation matrix reduces to the T2bCM defined in
Eq.~\eqref{eq:T2bCMgen} for $t\ge 0$. Like the T2bCM, the
time-ordered two-body correlation matrix is Hermitian,
positive-semidefinite, and also has a time-independent
trace. But unlike the T2bCM, the time-ordered two-body
correlation matrix is symmetric under combined time
reversal $T$ and index permutation $P$,
\begin{align} \label{eq:PTsymmetry}
T P\rho_2^{\mathrm{T}}(x,y; x',y';t)P T &\equiv
\rho_2^{\mathrm{T}}(y, x; y', x'; -t) =
\rho_2^{\mathrm{T}}(x,y; x',y';t) \,\, .
\end{align}
\end{widetext}
Using the spectral decomposition (\ref{eq:T2bCMdiag}) in
the general definition (\ref{eq:TO2bCM}), we obtain the
eigenvalues and corresponding eigenfunctions of the
time-ordered two-body correlation matrix in terms of
those of the T2bCM as
\begin{subequations}
\begin{align}
n_\alpha^\mathrm{T}(t) &= \theta(t)\, n_\alpha(t)
+ \theta(-t)\, n_\alpha(-t) \,\, , \\[0.1cm]
\chi^\mathrm{T}_\alpha(x, y; t) &= \theta(t)\,
\chi_\alpha(x, y; t) \pm \theta(-t)\, \chi_\alpha(y, x;
-t) \,\, .
\end{align}
\end{subequations}
Thus it is found that the eigenvalues
$n_\alpha^\mathrm{T}(t)$ of the time-ordered two-body
correlation matrix are even functions of $t$, and the
eigenfunctions are either even or odd under the $PT$
operation,
\begin{align} \label{eq:PTeigenf}
PT\chi^\mathrm{T}_\alpha(x, y; t) \equiv
\chi^\mathrm{T}_\alpha(y, x; -t) = \pm
\chi^\mathrm{T}_\alpha(x, y; t) \,\, .
\end{align}
The order parameter $\phi_0^\mathrm{T}(x, y; t) =
\sqrt{n_0^\mathrm{T}(t)}\,\,\chi_0^\mathrm{T}(x, y; t)$
associated with a macroscopic eigenvalue $n_0^\mathrm{T}
(t)$ then exhibits the same (anti)symmetry under $PT$ as
$\chi_0^\mathrm{T}(x, y; t)$.

Using $\phi_0^\mathrm{T}(x, y;t)$ as the order parameter
to describe pair condensation allows to draw parallels
to the formalism of Gor'kov anomalous Greens
functions~\cite{Gorkov1958} in terms of which
odd-frequency superconductivity was previously
discussed~\cite{Berezinskii1974,Linder2019}. The analogy
is most direct for the case when $n_0^\mathrm{T}(0)>0$
and $\chi_0^\mathrm{T}(x, y; t)$ is therefore odd under
$PT$~\footnote{This follows because, in the $t=0$ limit,
Eq.~\eqref{eq:PTeigenf} specialises to the relation
$P\chi^\mathrm{T}_\alpha(x, y; 0) = \pm
\chi^\mathrm{T}_\alpha(x, y; 0)$. Hence, odd-ness under
$P$ required of $\chi^\mathrm{T}_0(x, y; 0)\equiv\chi_0
(x, y; 0)$ when $n_0(0)>0$ implies odd-ness of
$\chi^\mathrm{T}_0(x, y; t)$ under $PT$.}. Then the
odd-in-$t$ part of the order parameter must be symmetric
under $P$, i.e., coincides with $S\phi_0^\mathrm{T}(x, y;
t)$, and odd-frequency pairing coexists with
even-frequency superfluidity described by
$A\phi_0^\mathrm{T}(x, y; t)$. It is this case (termed
the \emph{transformer scenario} in
Ref.~\cite{Thompson2024}) that is relevant to our present
study~\footnote{The situation when $n_0(0) = 0$
(considered as the \emph{generator scenario} in
Ref.~\cite{Thompson2024}) is generally more subtle and
does not pertain to the work presented here.}.

Here we investigate odd-frequency pairing in the
fermionic Tonks-Girardeau (FTG) gas by considering the
T2bCM \eqref{eq:T2bCMgen} and its dominant eigenvalue and
eigenfunction for $t \ge 0$, where the results apply
equally to the time-ordered two-body correlation matrix
\eqref{eq:TO2bCM}. The FTG gas is one of the rare
instances of an exactly solvable model for a system of
strongly interacting particles~\cite{Girardeau2004}. The
idea behind the model is strikingly simple: start with a
set of $N$ noninteracting bosonic quantum particles under
the influence of a single-particle Hamiltonian in one
spatial dimension,
\begin{align} \label{eq:Hsp}
H_\mathrm{sp} = -\frac{\hbar^2}{2 m} \, \frac{d^2}{d x^2}
+ V(x)\,\,  ,
\end{align}
where $V(x)$ is an arbitrary time-independent potential.
Then apply the Bose-Fermi mapping \cite{Girardeau1965} to
the many-body wave function to obtain the appropriate
fermionic wave function of the FTG gas. The resulting
fermionic model has infinitely strong attractive contact
interactions represented by a limit of the \v{S}eba
operator \cite{Seba1986a,Cheon1999,Granger2004,
Granet2022a}. 

The Bose-Fermi mapping is an identity in the
coordinate-space sector of ordered single-particle
coordinates $x_1 < x_2< \cdots <x_N$ and extends to the
full coordinate space by requiring fermionic antisymmetry
of the wave function. The Bose-Fermi mapping is
equivalent to the Jordan-Wigner transformation
\cite{Jordan1928} in lattice systems and is isospectral,
i.e., the energies and degeneracies of eigenstates are
the same for the FTG gas and the noninteracting Bose
gas~\cite{Yukalov2005}. Off-diagonal correlation
functions are not invariant under the Bose-Fermi mapping
and their determination can be nontrivial~\cite{Hao2007}.

The possibility to determine ground-state properties of
the FTG gas using its relation with a noninteracting Bose
gas \cite{Cheon1999,Girardeau2004,Granger2004} has
previously been employed to calculate
one-body~\cite{Girardeau2005,Bender2005,Hao2007,
Koscik2023,Sabater2024} and two-body~\cite{Girardeau2006,
Minguzzi2006} reduced density matrices, revealing the
signature features of macroscopic quantum order, i.e., a
dominant eigenvalue of the two-body reduced density
matrix that is proportional to the particle number $N$
\cite{Yang1962,Leggett2006}. Here we harness the same
Bose-Fermi duality for deriving an exact closed-form expression of the
T2bCM defined in Eq.~(\ref{eq:T2bCMgen}), enabling a
detailed study of macroscopic pairing order present in
the FTG gas. We present results for the
macroscopic eigenvalue $n_0(t)$ and the symmetric part
$S\chi_0(x, y; t)$ of the associated eigenfunction. It
is these two quantities that enter, via the relationship
\eqref{eq:TO2bCM} between the T2bCM and the time-ordered
two-body correlation matrix, into the part
$S\phi_0^\mathrm{T}(x, y; t)$ of the FTG-gas order
parameter that embodies odd-in-time pair correlations.
Throughout this work, we juxtapose results obtained for
the untrapped, homogeneous FTG gas and for the FTG gas
subject to a harmonic-oscillator confinement.

The remainder of this Article is organized as follows.
Section~\ref{sec:T2bCM} focuses on the FTG-gas T2bCM
whose exact general expression is presented and analyzed.
Results for the macroscopic T2bCM eigenvalue $n_0(t)$ and
an approximate form of the corresponding eigenfunction
$\chi_0(t)$ are obtained in Sec.~\ref{sec:ODLRO}. The
properties of the symmetric-pairing order parameter
$S\phi_0(x, y; t)$ are elucidated in
Sec.~\ref{sec:symmPair}. Section~\ref{sec:concl} presents
our conclusions. Details of longer mathematical
derivations and additional results are presented in
Appendices.

\section{Time-dependent two-body correlation matrix of
the FTG gas}
\label{sec:T2bCM}

\begin{figure*}
\includegraphics[height=3.7cm]{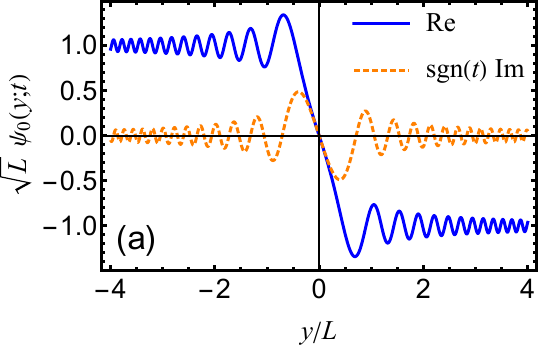}
\hfill\includegraphics[height=3.7cm]{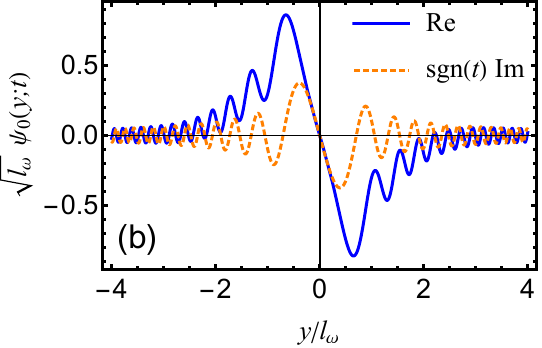}
\hfill\includegraphics[height=3.7cm]{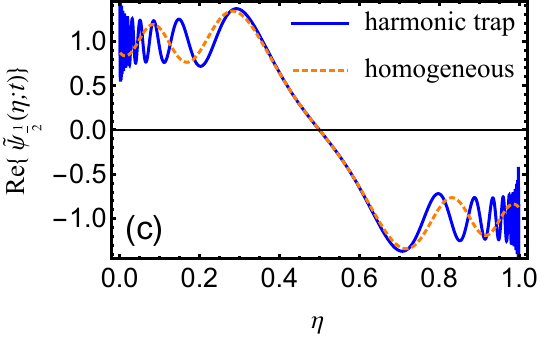}
\caption[]{\label{fig:Psi}%
Coordinate dependence of the dressed single-fermion wave
function $\psi_x(y; t)$ defined in Eq.~(\ref{eq:PsiDef})
for (a)~the homogeneous FTG gas with $\omega_L t = 0.1$
and (b)~the FTG gas in a harmonic-oscillator trap with
$\omega t = 0.1$. Here the real and imaginary parts are
plotted as the solid blue and dashed orange curves,
respectively. The imaginary part switches sign when $t\to
-t$.  Panel~(c) plots the dependence of the real part of
the unit-less dressed single-fermion wave function on the
universal coordinate $\eta\equiv F(y)$ [see
Eqs.~(\ref{eq:Fdef}) and (\ref{eq:transPsiDef})] for both
the homogeneous and the harmonically trapped FTG gases
with $\omega_L t = \omega t/\pi = 0.01$.}
\end{figure*}

The FTG gas is a system of strongly interacting fermions,
potentially subject to an external trapping potential
$V(x)$. Despite being highly correlated, its $N$-particle
ground state $\ket{\Psi_0}$ is known exactly from
application of the Bose-Fermi mapping
\cite{Girardeau1965,Cheon1999,Granger2004,Girardeau2004}.
Using the same mapping, it is also possible to derive a
closed-form expression for the T2bCM that generalises an
earlier result for the two-body reduced density matrix
\cite{Girardeau2006,Minguzzi2006} to include a $t$
dependence (see Appendix~\ref{app:FBmappCalc} for
details). The expression, which will form the basis of
all further analysis in this Article, is explicit in
terms of integrals that can be evaluated numerically. It
reads
\begin{align}\label{eq:T2bCM1st}
& \rho_2(x, y ;x', y'; t) = N(N-1)\,\phi^*(x)\,
\psi^*_x(y; t) \nonumber \\[0.1cm]
& \hspace{1.5cm} \times \phi(x')\,\psi_{x'}(y'; t)\,
\left[ P(x, y; x', y';t) \right]^{N-2} \,\, .
\end{align}
Here $\phi(x)$ is the lowest-energy eigenstate of the
single-particle Hamiltonian of Eq.~\eqref{eq:Hsp}, with
its eigenvalue denoted by $E_0$. We introduced the
$t$-dependent functions
\begin{align}\label{eq:PsiDef}
\psi_x(y; t) = e^{-i\frac{E_0}{\hbar} t} \int dz\,\,
G(y; z; -t)\, \sgn(x- z)\, \phi(z)
\end{align}
and
\begin{align}\label{eq:PfacDef}
& P(x, y; x', y'; t) = \nonumber \\[0.1cm]
& \hspace{0.5cm} \int dz\,\, \sgn(y - z)\, \sgn(y' - z)
\, \psi^*_x(z; t)\, \psi_{x'}(z; t) \,\, .
\end{align}
As the FTG gas is a model for a system of fermions in one
spatial dimension, all coordinates $x$, $y$, $z$ etc.\
refer to this single dimension.

The fundamental $t$-dependent building block in the
T2bCM expression (\ref{eq:T2bCM1st}) is $\psi_x(y; t)$
whose definition involves the single-particle propagator
[see Eq.~(\ref{eq:PsiDef})]
\begin{align}
G(y; z; t) = \Big\langle y\Big|\exp\Big(-i\,
\frac{t}{\hbar} \, H_\mathrm{sp}\Big)\Big|z\Big\rangle
\,\, .
\end{align}
We refer to $\psi_x(y; t)$ as the \emph{dressed
single-fermion wave function}. The motivation for this
terminology is easiest to understand in the $t=0$ limit.
Note that the product
\begin{align}
\phi(x)\psi_x(y; 0) =\phi(x)\, \sgn(x-y)\, \phi(y)
\end{align}
is the ground-state eigenfunction of a two-particle FTG
system. As discussed before in Ref.~\cite{Cheon1999},
it has the appropriate fermionic antisymmetry under $x
\leftrightarrow y$ exchange and a discontinuity at
$x=y$, as characteristic for the FTG gas. The part
$\psi_x(y; 0) = \sgn(x-y)\, \phi(y)$ corresponds to the
single-particle eigenfunction $\phi(y)$ dressed with the
statistical phase $\sgn(x-y)$ from the relative motion of
two fermions. Thus, $\psi_x(y; 0)$ represents the quantum
amplitude of a fermion subject to the trap potential
$V(y)$ in the presence of another fermion located at
point $x$. Evolving $\psi_x(y; 0)$ under the
single-particle Hamiltonian over $-t$ yields $e^{i E_0
t/\hbar}\,\psi_x(y;t)$, which can be interpreted as the
time-evolved wave function of the remaining particle
after one particle was removed from the two-particle
ground state at position $x$ and time
$t=0$~\footnote{The negative sign for the $t$-evolution
interval just reflects our convention for defining the
T2bCM \eqref{eq:T2bCMgen} and has no special
significance.}. The $E_0$-dependent phase factor
representing the trivial time evolution of the eigenstate
$\phi(y)$ is of no physical importance and has been
removed in our definition (\ref{eq:PsiDef}) of
$\psi_x(y;t)$.

Using the explicit form of the propagator for a free
particle~\cite{Sakurai2011}, i.e., for $V(y)=0$, we
obtain
\begin{align}\label{eq:FreeGasPsi}
\psi_x(y; t) &= \phi(y)\,\,\mathrm{erf}\left(
\sqrt{\frac{i}{2\omega_L t}}\,\frac{x - y}{L}\right)\, ,
\end{align}
where $L$ denotes the system size, $\phi(y) = 1/\sqrt{L}$,
and $\omega_L = \hbar/(m L^2)$. Assuming instead a
parabolic confinement $V(y) = m\omega^2 y^2/2$ and
utilising the known harmonic-oscillator
propagator~\cite{Sakurai2011,Holstein1998,Thornber1998,
Chaos1999}, we find
\begin{align}\label{eq:HarmTrapPsi}
\psi_x(y; t) &= \phi(y)\,\erf\left[\sqrt{\frac{i}{2\sin
(\omega t)}}\,\frac{e^{-i \frac{\omega t}{2}}\, x
- e^{i\frac{\omega t}{2}}\, y}{l_\omega}\right] \, ,
\end{align}
with the oscillator length $l_\omega = \sqrt{\hbar/(m
\omega)}$ and $\phi(y) = \exp[-y^2/(2\,l_\omega^2)]/
\sqrt{\sqrt{\pi}\,l_\omega}$. See
Appendix~\ref{app:dressedWF} for details of derivations
and additional results. The coordinate dependence of the
dressed single-fermion wave functions of
Eqs.~(\ref{eq:FreeGasPsi}) and (\ref{eq:HarmTrapPsi}) is
plotted, respectively, in panels (a) and (b) of
Fig.~\ref{fig:Psi}.

Direct comparison between FTG gases subject to different
external potentials is facilitated by a particular
variable transformation. It was shown in
Ref.~\cite{Koscik2023} that the reduced density matrices
of the FTG ground state can be brought into a universal
form, with the implication that their eigenvalues are
independent of the shape of the trapping potential
$V(x)$. The transformation proceeds by replacing the
position variable by the cumulative
probability-distribution function of the single-particle
ground state,
\begin{align}\label{eq:Fdef}
F(y) = \int_{-\infty}^y dz\,\, |\phi(z)|^2 \,\, .
\end{align}
Defining $\xi=F(x)$ and $\eta=F(y)$, as well as
\begin{subequations}
\begin{align}\label{eq:transGdef}
G(y; z; t) &= \phi(y)\, \phi^*(z)\, e^{i\frac{E_0}{\hbar}
t}\, \tilde{G}(F(y); F(z); t) \,\, , \\[0.1cm]
\label{eq:transPsiDef}
\psi_x(y; t) &= \phi(y)\,\, \tilde{\psi}_{F(x)}(F(y); t)
\,\, ,
\end{align}
\end{subequations}
a completely unit-less dressed single-fermion wave
function $\tilde{\psi}_\xi(\eta; t)\equiv
\tilde{\psi}_{F(x)}(F(y); t)$ emerges as
\begin{align}\label{eq:UniPsi}
\tilde{\psi}_\xi(\eta; t) &= \int_0^1 d\zeta \,\,
\tilde{G}(\eta; \zeta; -t)\,\, \sgn(\xi - \zeta) \,\, .
\end{align}
Thus the transformation (\ref{eq:Fdef}) eliminates form
factors related to the single-particle eigenstate
$\phi(z)$ so that the only influence of the external
potential is on the $t$ evolution transpired by the
propagator $\tilde{G}(\eta; \zeta; -t)$. In contrast to
the ($t$-independent) two-body reduced density matrix and
its eigenfunctions, which become completely universal by
these variable transformations, the transformed dressed
single-fermion wave function retains a residual
dependence on the external potential. See panel~(c) of
Fig.~\ref{fig:Psi} for illustration. The line shapes
corresponding to the two systems' bulk regions coincide
for $\omega_L t = \omega t/\pi$. The residual differences
exhibited near the system boundaries are significant for
the discussion of odd-frequency pairing correlations; see
below.

\section{Macroscopic T2bCM eigenvalue and ODLRO in the
FTG gas}
\label{sec:ODLRO}

ODLRO is expressed by the, at least approximate,
factorization of the T2bCM when its spectral
representation (\ref{eq:T2bCMdiag}) is dominated by the
term associated with a macroscopic eigenvalue $n_0(t)
\sim N$;
\begin{align}
\rho_2(x, y; x', y'; t) \approx n_0(t) \big[
\chi_0(x, y; t)\big]^*\, \chi_0(x', y'; t)\,\, .
\end{align}
From the general expression (\ref{eq:T2bCM1st}), it is
obvious that the T2bCM is trivially factorized for $N=2$,
with $\chi_0(x, y; t) = \phi(x)\,\psi_x(y; t)$ and the
eigenvalue $n_0(t)=2$ being independent of $t$. Finding
the T2bCM's spectral decomposition for $N>2$ is
complicated due to the nontrivial structure of the
function $P(x, y; x', y'; t)$ defined by
Eq.~(\ref{eq:PfacDef}). See the general form
(\ref{eq:T2bCM1st}) of the T2bCM.

Progress can be made with an off-diagonal long-range
approximation inspired by the ODLRO argument of
Ref.~\cite{Yang1962} and previously employed for the FTG
gas in Ref.~\cite{Girardeau2006}, where the pair $x$ and
$y$ of coordinates are well-separated from the pair $x'$
and $y'$, but the intra-pair distance is much smaller
than the inter-pair separation. Under these conditions,
the matrix elements of the two-body reduced density
matrix of a normal Fermi gas decay exponentially, while
a macroscopic Cooper-pair condensate provides a
contribution that remains finite~\cite{Yang1962}. At
$t>0$, particle locations get smeared out due to
wave-packet spreading that occurs on the $t$-dependent
length scale $\ell_t=\sqrt{\hbar |t|/m}$. For this to
not matter, we have to also assume that the members of
a pair are separated far apart compared to $\ell_t$.
More formally, the off-diagonal long-range approximation
can be summarized as
\begin{equation}\label{eq:genODLRO}
\ell_t\ll |x-y| \approx |x'-y'| \ll |x + y -(x'+y')|
\approx \mbox{system size}\, .
\end{equation}

Under the conditions specified in
Eq.~(\ref{eq:genODLRO}), we find~\cite{Wong2026}
\begin{equation}\label{eq:PtoSmallP}
P(x, y; x', y'; t)\approx 1 - p^*(x, y; t) - p(x', y'; t)
\,\, ,
\end{equation}
with $0\le |p(x, y; t)| \ll 1$. This makes it possible
to further approximate~\cite{misc:PexpoApprox}
\begin{equation}\label{eq:PexpoN}
\left[P(x, y; x', y'; t)\right]^{N-2} \approx e^{-(N-2)
\, p^*(x, y; t)}\,\, e^{-(N-2)\, p(x', y'; t)} \, .
\end{equation}
One can then obtain the normalized macroscopic
eigenfunction of the T2bCM in approximated form as
\begin{align}\label{eq:chiNgen}
\chi_0(x, y; t) &= \mathcal{C}_t\, \phi(x)\,\psi_x(y;t)
\,\, e^{-(N-2)\, p(x, y; t)} \,\, ,
\end{align}
corresponding to the eigenvalue
\begin{align}\label{eq:ODLROeval}
n_0(t) &= N(N-1)\, \left|\mathcal{C}_t\right|^{-2}\,\, . 
\end{align}
Upto a trivial global phase, the normalization
factor $\mathcal{C}_t$ is defined by the condition $\int
|\chi_0(x, y; t)|^2\, dx\, dy = 1$ and can be determined
via
\begin{subequations}
\begin{align}\label{eq:genNorm}
&\left|\mathcal{C}_t\right|^{-2} \nonumber \\[0.1cm]
&= \int dx\, |\phi(x)|^2 \int dy \, |\psi_x(y;t)|^2 \,
e^{-2(N-2)\Re\{p(x, y; t)\}}\,\, , \\ \label{eq:uniNorm}
&=\int_0^1 d\xi\, \int_0^1 d\eta \,\, \big|
\tilde{\psi}_\xi(\eta; t) \big|^2 \, e^{-2(N-2)
\Re\{\tilde{p}(\xi,\eta; t)\} } \,\, .
\end{align}
\end{subequations}
The form (\ref{eq:uniNorm}) utilizes the universal
coordinates $\xi=F(x)$ and $\eta=F(y)$ [see
Eq.~(\ref{eq:Fdef})] and involves $\tilde{p}(\xi, \eta;
t)\equiv \tilde{p}(F(x), F(y);t) = p(x, y; t)$, as well
as $\tilde{\psi}_\xi(\eta; t)$ [Eq.~(\ref{eq:UniPsi})].

The $t=0$ limit can be discussed explicitly. Evaluating
Eq.~(\ref{eq:uniNorm}) with the universal expressions
$\tilde{\psi}_\xi(\eta; 0) = \sgn(\xi - \eta)$ and
$\tilde{p}(\xi, \eta; 0) = 2|\xi-\eta|$, we find
\begin{align}
\left|\mathcal{C}_0\right|^{-2} &= \frac{1}{2(N-2)} -
\frac{1}{8(N-2)^2}\left[ 1 - e^{-4(N-2)} \right] \,\, .
\end{align}
The largest two-body reduced density matrix eigenvalue is
thus obtained as
\begin{align}\label{eq:uni2bRDMeval}
n_0(0) = \frac{N(N-1)}{2(N-2)} - \frac{N(N-1)}{8(N-
2)^2}\left[ 1 - e^{-4(N-2)} \right] \,\, .
\end{align}
The analytical result (\ref{eq:uni2bRDMeval}) agrees well
with previous numerical work~\cite{Minguzzi2006} and is
completely independent of the presence or shape of the
trapping potential $V(x)$, in accordance with
Ref.~\cite{Koscik2023}. For large particle number $N \to
\infty$, we obtain $n_0(0)\to N/2$, which means that the
FTG ground state is a macroscopic Cooper-pair condensate
for any trapping potential. In an extended system with a
thermodynamic limit, the FTG gas has true
ODLRO~\cite{Brand2026}.

\begin{figure}[t]
\includegraphics[width=0.45\textwidth]{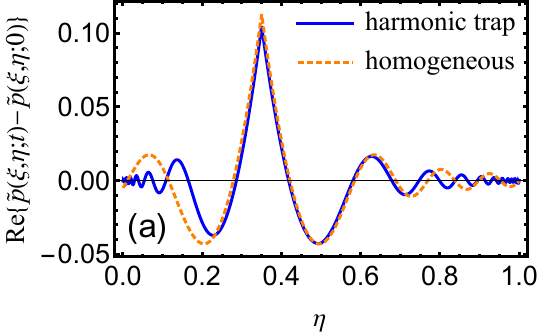}\\[0.5cm]
\includegraphics[width=0.45\textwidth]{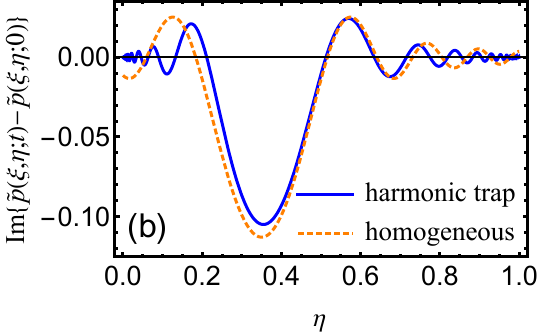}
\caption[]{\label{fig:pfunc}%
Comparison of the purely $t$-dependent contribution to
$\tilde{p}(\xi,\eta;t)$ for the homogeneous and the
harmonically trapped FTG gases with $\omega_L t = \omega
t/\pi = 0.01$ and $\xi=0.35$.}
\end{figure}

Calculation of the $t$-dependent eigenvalue $n_0(t)$
proceeds from $p(x, y; t)$ obtained~\cite{Wong2026},
respectively, for the homogeneous FTG gas,
\begin{align}\label{eq:FreeGasLitP}
p_\mathrm{hom}(x, y; t) &= \frac{|x-y|}{L}\left[ 2 +
\frac{1}{\sqrt{\pi}}\, \Gamma\left( -\frac{1}{2},
\frac{i(x-y)^2}{2\omega_L t L^2} \right)\right] ,
\end{align}
and for the FTG gas confined by a harmonic-oscillator
potential,
\begin{align}\label{eq:HarmTrapLitP}
p_\mathrm{osc}(x, y; t) &= \left| \mathrm{erf}\left(
\frac{x}{l_\omega} \right) - \mathrm{erf}\left(
\frac{y}{l_\omega} \right)\right| \nonumber \\[0.1cm]
&\hspace{0.2cm}+
\frac{e^{-\frac{|x\, y|}{l_\omega^2}}}{\pi}\,
\frac{|x-y|}{l_\omega}\,\Gamma\left( -\frac{1}{2},
\frac{i(x-y)^2}{2\omega t\, l_\omega^2} \right) .
\end{align}
Here $\Gamma(a, z)$ denotes the incomplete Gamma
function  discussed, e.g., in Sec.~6.5 of
Ref.~\cite{Abramowitz1972}. The $t=0$ limit of these
expressions recovers previous
results~\cite{Girardeau2006,Minguzzi2006}, and their
$t$-dependent parts are compared using universal
coordinates $\xi=F(x)$ and $\eta=F(y)$ in
Fig.~\ref{fig:pfunc}. An approximate coincidence in the
region $\xi\approx\eta$ occurs again for $\omega_L t =
\omega t/\pi$ and $\xi,\eta$ not too close to the
boundary.

\begin{figure}[t]
\includegraphics[width=0.45\textwidth]{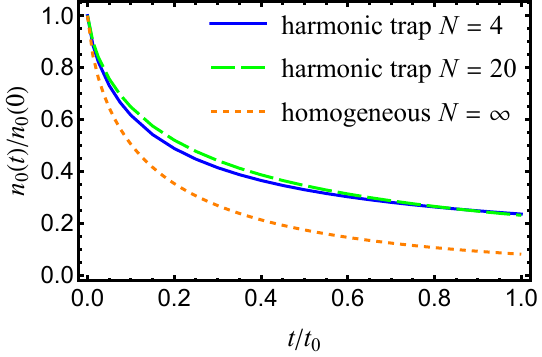}
\caption[]{\label{fig:evalComp}%
Dependence of the dominant T2bCM eigenvalue on $t>0$. The
solid blue (dashed green) curve is for $N=4$ ($N=20$)
in a harmonic trap and $t_0=\pi/[(N-2)^2\omega]$. The
dotted orange curve plots the thermodynamic limit
(\ref{eq:evalTdy}).}
\end{figure}

The expression (\ref{eq:genNorm}) is straightforwardly
evaluated using (\ref{eq:FreeGasPsi}) and
(\ref{eq:FreeGasLitP}), yielding the $t$-dependent
macroscopic eigenvalue for the homogeneous FTG gas as
\begin{align}\label{eq:FreeGasN0}
n_0(t) &= \frac{N(N-1)}{2(N-2)}\,\Theta_1\left(
\frac{t}{t_0}\right) \nonumber \\[0.1cm] &\hspace{0.5cm}
-\frac{N(N-1)}{8(N-2)^2}\left[\Theta_2\left(\frac{t}{t_0}
\right)-\Theta_3^{(N)}\left(\frac{t}{t_0}\right)\right] ,
\end{align}
with the time scale
\begin{align}\label{eq:t0Hom}
t_0 &= \frac{m}{\hbar} \left(\frac{L}{N-2}\right)^2\,\, .
\end{align}
The $N$-independent functions $\Theta_{1,2}$ as well as
$\Theta_3^{(N)}$ are known in terms of definite
integrals; see Appendix~\ref{app:ThetaFuncs} for
details. As the expression (\ref{eq:FreeGasN0}) is
derived using the off-diagonal long-range approximation
of Eq.~(\ref{eq:genODLRO}), it is only valid for $|t|<
\omega_L^{-1}$. Taking the usual homogeneous-system
thermodynamic limit $N\to\infty$, $L\to\infty$, $N/L
\equiv\rho_0=\,$const., we obtain
\begin{align}\label{eq:evalTdy}
n_0(t)\to \frac{N}{2}\, \Theta_1\left(\frac{t}{t_0}
\right) \,\, , \mbox{ with }\,\, t_0\to \frac{m}{\hbar
\rho_0^2} \,\, .
\end{align}
Thus macroscopic order is maintained at finite $t$, as
signified by the asymptotic proportionality of $n_0(t)$
and $N$, with a prefactor that is monotonously reduced as
$t$ increases. See the plot of the relation
\eqref{eq:evalTdy} in Fig.~\ref{fig:evalComp}.

\begin{figure*}[t]
\includegraphics[height=4cm]{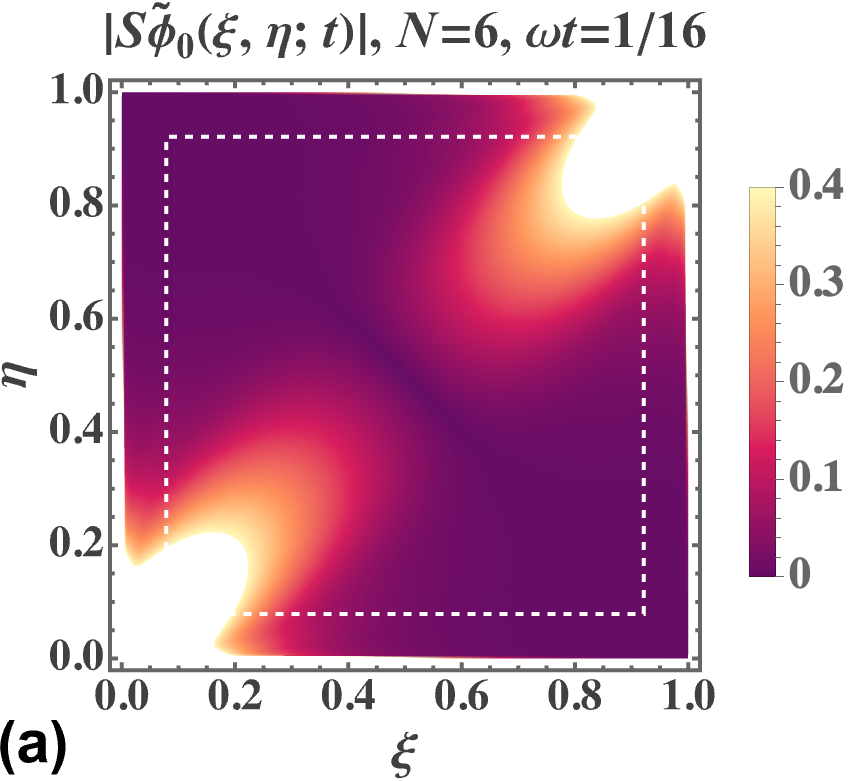}
\hfill\includegraphics[height=4cm]{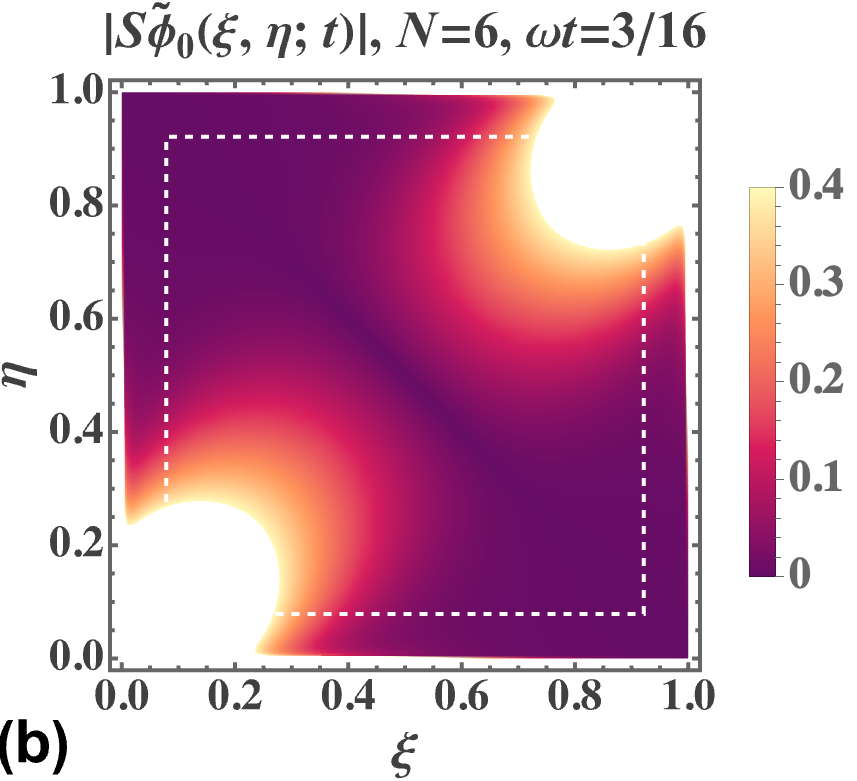}
\hfill\includegraphics[height=4cm]{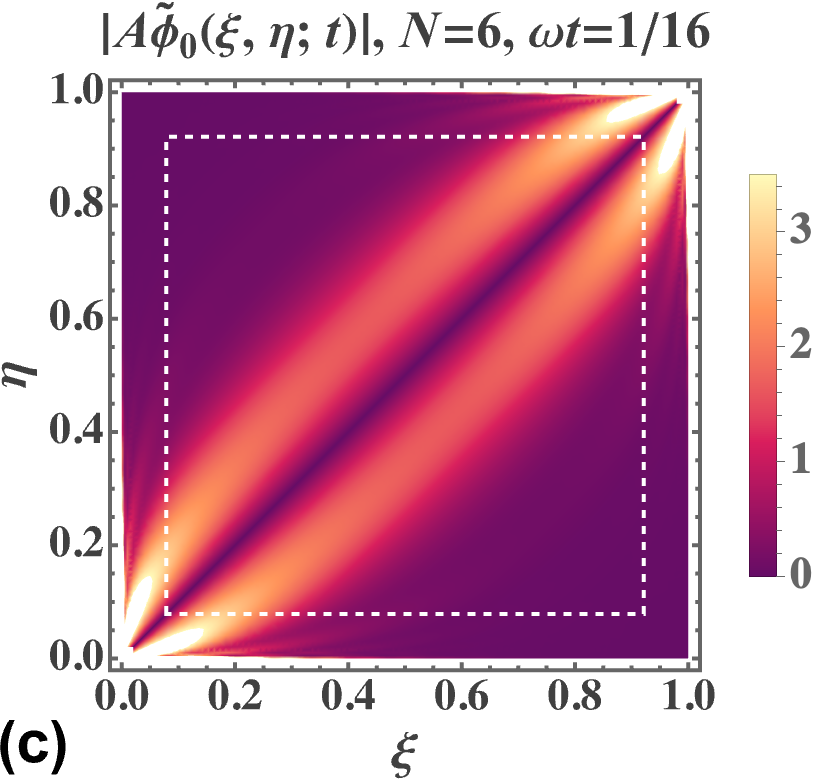}
\hfill\includegraphics[height=4cm]{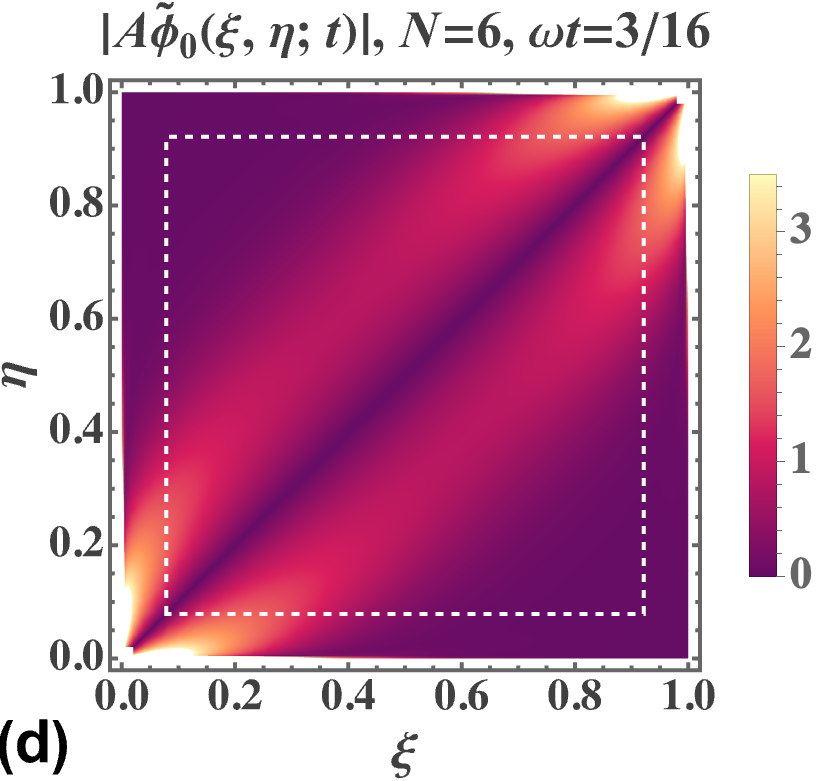}\\[0.1cm]
\includegraphics[height=4cm]{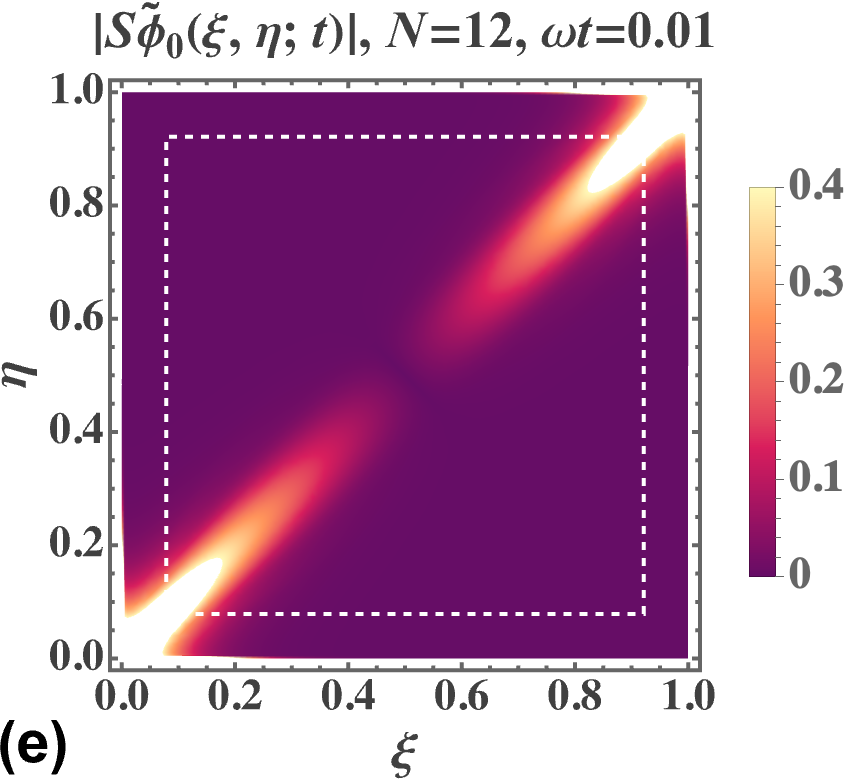}
\hfill\includegraphics[height=4cm]{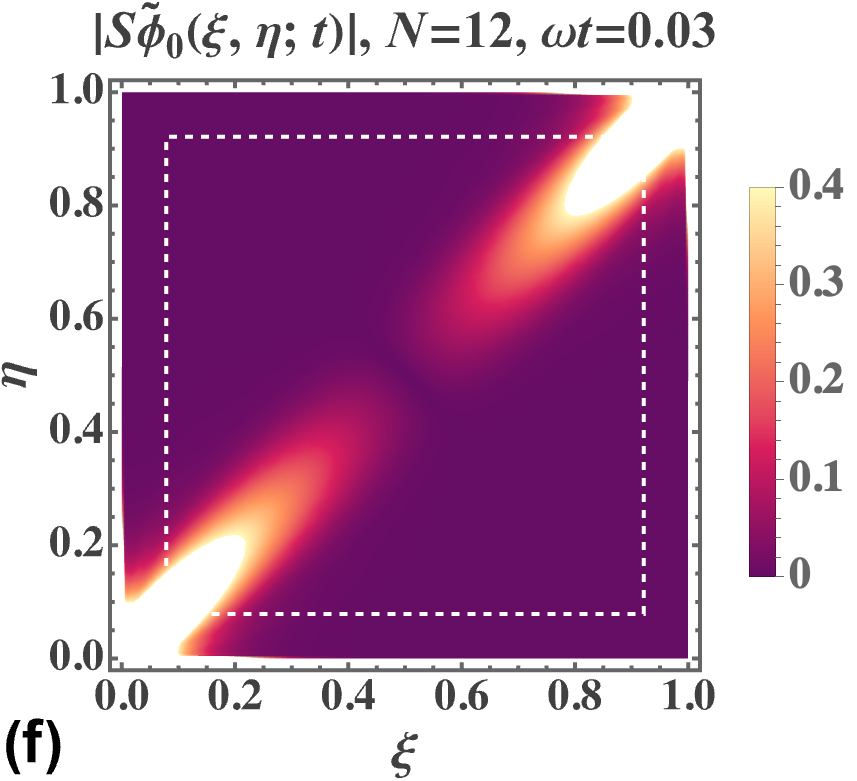}
\hfill\includegraphics[height=4cm]{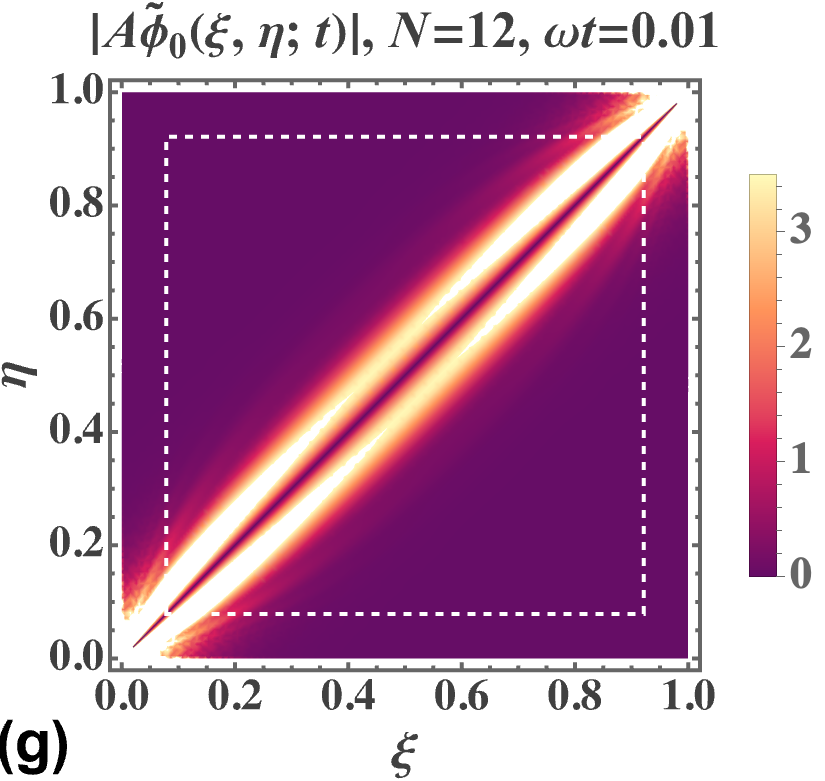}
\hfill\includegraphics[height=4cm]{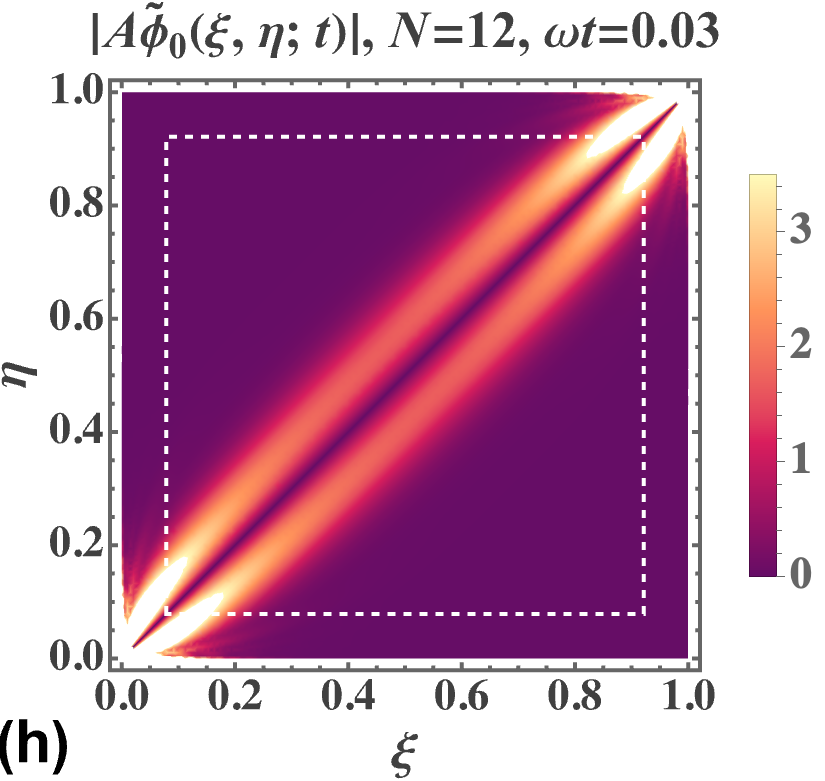}
\caption[]{\label{fig:ODLROop}%
Pair-condensate order parameter for a harmonically
trapped FTG gas, calculated from Eq.~(\ref{eq:phiNgen}).
The modulus of the symmetric [antisymmetric] part is
plotted in panels (a), (b), (e) and (f) [(c), (d), (g)
and (h)] using universal coordinates $\xi=F(x)$ and
$\eta=F(y)$. The upper (lower) row shows results for
$N=6$ ($N=12$) and two values of $t$ differing by a
factor of $3$. Dashed rectangles delimit the region
$|x|,|y|\le l_\omega$ where the FTG gas is largely
concentrated.}
\end{figure*}

The $t$ and $N$ dependences of the largest T2bCM
eigenvalue for the harmonically trapped FTG gas have not
been obtained in analytical closed form, but they can be
accessed numerically. The qualitative behavior is quite
similar to the homogeneous FTG gas. To enable a direct
comparison, we define the characteristic time scale $t_0
=\pi/[(N-2)^2 \omega]$ that, in the thermodynamic limit
for a parabolic potential~\cite{Minguzzi2006} ($N\to
\infty$, $\omega\to 0$, $N^2\omega \equiv\pi\hbar\rho_0^2
/m=\,$const.) yields again $t_0\to m/(\hbar \rho_0^2)$.
The largest-eigenvalue renormalization at finite $t$ in a
harmonic trap is illustrated in Fig.~\ref{fig:evalComp}.
A relation analogous to Eq.~\eqref{eq:evalTdy} is
evidenced by the numerical results, indicating
proportionality of $n_0(t)$ and $N$ for large $N$. Thus
macroscopic order is maintained at finite $t$ also in the
harmonically trapped FTG gas.

\section{Odd-frequency pair correlations emerging at the
FTG-gas boundary}
\label{sec:symmPair}

Odd-in-time, or odd-frequency pairing is embodied in the
part $S\phi_0(x, y; t)$ of the order parameter that
is symmetric under the exchange of particle coordinates
\cite{Linder2019,Thompson2024}. See
Eq.~(\ref{eq:symmPhi}) for the definition. Combining the
results from Eqs.~\eqref{eq:chiNgen} and
\eqref{eq:ODLROeval} in the expression \eqref{eq:OPdef}
yields
\begin{align}\label{eq:phiNgen}
\phi_0(x, y; t) &= \sqrt{N(N-1)}\,\, \phi(x)\,\psi_x(y;t)
\,\, e^{-(N-2)\, p(x, y; t)}
\end{align}
as the FTG-gas order parameter. Its (anti)symmetric
contribution is determined by the (anti)symmetric part of
$\phi(x)\,\psi_x(y; t)$  since the function $p(x,y;t)$
turns out to be  symmetric under exchange of $x$ and $y$
for the situations considered in this work; see
Eqs.~(\ref{eq:FreeGasLitP}) and (\ref{eq:HarmTrapLitP}).

For the homogeneous FTG gas, we find $\phi(x)\,\psi_x(y;
t)=-\phi(y)\,\psi_y(x; t)$ at any $t$ [see
Eq.~(\ref{eq:FreeGasPsi})]. This system has therefore a
fully antisymmetric order parameter; $S\phi_0(x, y;
t)\equiv 0$ even for $t>0$, and no odd-frequency pairing
correlations are exhibited. 

The harmonically trapped FTG gas develops a symmetric
contribution to the pair-condensate order parameter at
finite $t$. We plot the symmetric and antisymmetric parts
of $\phi_0(x, y; t)$ for this case in
Fig.~\ref{fig:ODLROop}. A striking feature is the strong
spatial concentration of $S \phi_0(x, y; t)$ near the
system boundaries $x\approx y\sim \pm l_\omega$. In
contrast, $A \phi_0(x, y; t)$ is seen to extend
throughout the system's bulk. The vanishing of
symmetric-pairing correlations in the homogeneous system
and its boundary localization in the presence of a trap
suggests that odd-frequency pairing in the FTG gas is a
mesoscopic edge effect.

\begin{figure}[b]
\includegraphics[width=\columnwidth]{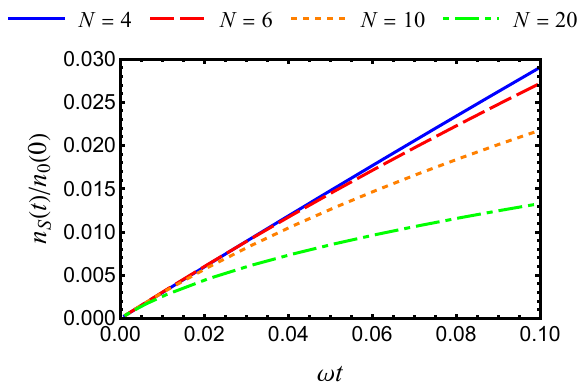}
\caption[]{\label{fig:nS(t)}%
Magnitude of the odd-frequency, symmetric-pairing part of
the FTG-gas superfluid order parameter. The quantity
$n_S(t)$, defined in Eqs.~(\ref{eq:nS(t)def&alt}), is
plotted as a function of $t$ for several values of the
particle number $N$.}
\end{figure}

An overall quantification of odd-frequency pairing in the
FTG gas is provided by the squared norm of the symmetric
part of the order parameter given in
Eq.~(\ref{eq:phiNgen}),
\begin{subequations}\label{eq:nS(t)def&alt}
\begin{align}\label{eq:nS(t)def}
n_S(t) &= \int dx \int dy \,\, \left| S\phi_0(x, y; t)
\right|^2 \,\, , \\[0.1cm] \label{eq:nS(t)alt}
&= n_0(t) \int dx \int dy \,\, \left| S\chi_0(x, y; t)
\right|^2\,\, .
\end{align}
\end{subequations}
We plot the $t$ dependence of this quantity in
Fig.~\ref{fig:nS(t)}, using $n_0(0)$ as normalization.
Qualitatively, our results show that $n_S(t)/n_0(0)$ is
suppressed with increasing $N$ for any fixed $t$,
indicating that the odd-frequency, symmetric-pairing
correlations in the harmonically trapped FTG gas are not
representative of a macroscopic order.

As is apparent from Eq.~\eqref{eq:nS(t)alt}, $n_S(t)$ is
the product of $n_0(t)$ and the squared norm of $S\chi_0
(x, y; t)$. The interplay of the $t$ and $N$ dependences
from both factors gives rise to a complex behavior
exhibited by $n_S(t)$. Results from the previous
Sec.~\ref{sec:ODLRO} show that $n_0(t)$ decreases
monotonically from $n_0(0)$ for increasing $t$ (see
Fig.~\ref{fig:evalComp}). In contrast, the vanishing of
$S\chi_0(x, y; t)$ in the limit $t\to 0$ implies that its
norm will be an increasing function of $t$ in the
small-$t$ limit, but this increase is limited by the
overall normalization of $\chi_0(x, y; t)$. To
characterize the magnitude of odd-frequency pairing as a
function of relevant parameters in certain limits, we
postulate a scaling form
\begin{align}\label{eq:nS(t)scale}
n_S(t) \approx c\, N^a\, (\omega t)^b \,\, .
\end{align}
We observe that the \textit{Ansatz\/}
\eqref{eq:nS(t)scale} is obeyed in two different regimes
of $t$ separated by an $N$-dependent crossover scale
$t_\mathrm{c}(N)\propto (N-2)^{-2/3}\,\omega^{-1}$. A
scaling analysis was performed to extract universal
results for the exponents in the fixed-$t$, large-$N$
limit of the $t_\mathrm{c}(N) < t \ll \omega^{-1}$
regime, which is the one that is relevant for discussing
macroscopicity and long-range order at finite time $t$.
Details are given in Appendix~\ref{app:scaling}, leading
to the results $a\approx 0.05$, $b\approx 0.52$, and
$c\approx 0.48$. The scaling $n_S(t)\propto N^a$ with
$0\lesssim a\ll 1$ in the large-$N$ limit is again
indicative of the mesoscopic, nonextensive character of
odd-frequency pairing in the FTG gas. The value of $a$
being so close to zero suggests the absence of even
algebraic ODLRO~\cite{Coleman1993a,Colcelli2018}.

\section{Conclusions}
\label{sec:concl}

We generalize the concept of pairing order in the FTG
gas to include the macroscopic order parameter's
dependence on a finite time delay $t$ between members of
a fermion pair, opening up the possibility to investigate
odd-in-$t$ (also called odd-frequency) pair correlations
in a system with fixed particle number $N$. The formalism
is based on considering the largest eigenvalue
$n_0^\mathrm{T}(t)$ and associated eigenfunction
$\chi_0^\mathrm{T}(x, y; t)$ of the time-ordered two-body
correlation matrix \eqref{eq:TO2bCM}. The order parameter
$\phi_0^\mathrm{T}(x, y; t) = \sqrt{n_0^\mathrm{T}(t)}
\,\, \chi_0^\mathrm{T}(x, y; t)$ for the FTG gas is
antisymmetric under $PT$; i.e., fermion-pair index
exchange ($P$) combined with time inversion $t\to -t$
($T$). Thus it is analogous to the anomalous Gor'kov
Greens function~\cite{Gorkov1958} whose definition
requires indeterminate particle number and in terms of
which odd-frequency pairing has previously been
discussed~\cite{Berezinskii1974,Linder2019}. The part of
the order parameter $\phi_0^\mathrm{T}(x, y; t)$ that is
symmetric under $P$ therefore represents pair
correlations antisymmetric under $T$, i.e., odd-frequency
pairing.

We specifically present results for $t\ge 0$ where the
time-ordered two-body correlation matrix of
Eq.~\eqref{eq:TO2bCM} coincides with the time-dependent
two-body correlation matrix (T2bCM) of
Eq.~\eqref{eq:T2bCMgen} that was introduced as a general
theoretical concept in recent work~\cite{Thompson2024}.
The exact expression \eqref{eq:T2bCM1st} for the T2bCM
is obtained by virtue of the Bose-Fermi mapping
technique. Its approximate factorization is accomplished
within the off-diagonal long-range approximation [as per
conditions expressed in Eq.~\eqref{eq:genODLRO}], enabling
determination of its largest eigenvalue
\eqref{eq:ODLROeval} and the order parameter
\eqref{eq:phiNgen} in terms of functions $\psi_x(y;t)$
[defined in Eq.~\eqref{eq:PsiDef}] and $p(x, y; t)$
[implicitly defined in Eq.~\eqref{eq:PtoSmallP}]. We have
calculated these functions for both the homogeneous FTG
gas and the FTG gas trapped in a harmonic-oscillator
potential.

We find that macroscopicity of the largest T2bCM
eigenvalue persists at fixed finite $t$, i.e.,
$n_0(t)\propto N$ for $N\to\infty$. See
Fig.~\ref{fig:evalComp}. An analytical formula is
obtained for $n_0(0)$ [see Eq.~\eqref{eq:uni2bRDMeval}]
for which previously only numerical results were
available~\cite{Minguzzi2006}. The part $S\phi_0(x, y;
t)$ of the order parameter describing odd-frequency pair
correlations vanishes in the homogeneous FTG gas and
becomes significant only at the boundary of the
harmonically trapped system (see Fig.~\ref{fig:ODLROop}).
The magnitude of odd-frequency pairing, as defined in
Eqs.~(\ref{eq:nS(t)def&alt}), is nonextensive
thermodynamically (see Fig.~\ref{fig:nS(t)}).

The present work extends the formalism developed
earlier~\cite{Thompson2024} to describe odd-frequency
pairing in systems with fixed particle number. Previously
discussed examples~\cite{Thompson2024} included the
Zeeman-split Fermi superfluid and the composite-boson
condensate, both of which are scenarios where the
odd-frequency part of the order parameter is itself
macroscopic. The present study of the FTG gas reveals
mesoscopic odd-frequency pair correlations appearing in
nonhomogeneous situations such as in the presence of a
trap potential. This model can thus represent the class
of systems where odd-frequency pairing is tied to
boundaries or interfaces~\cite{Bergeret2005,Eschrig2007, 
Tanaka2007}, which is a promising experimental platform
for realizing this type of hidden
order~\cite{DiBernardo2015,Pal2017,Krieger2020,
Perrin2020}. Thus our work opens up new possibilities for
understanding odd-frequency pairing in the FTG gas and
related fixed-particle-number model systems
\cite{Brand2026} to inform realistic experimental studies.

Pair correlations of ultracold atoms can be measured
using quantum noise interferometry \cite{Altman2004a,
Folling2005a} or with quantum gas microscopes, which
achieve single-particle accuracy~\cite{Holten2022,
Yao2025,DeJongh2025,Mark2025}. To reveal odd-frequency
correlations, however, new protocols will need to be
developed to probe pair correlations with a time delay.
Ideas for doing so have been proposed for angle-resolved
photoelectron spectroscopy in the context of electronic
superconductivity~\cite{Kornich2021}. Adapting these
ideas for quantum gas microscopes, possibly combining
techiques of nondestructive imaging \cite{Andrews1996}
and high-resolution microscopy~\cite{Verstraten2025}
could facilitate a direct measurement of the T2bCM, thus
revealing local odd-frequency pair correlations.

\begin{acknowledgments}

K.T.\ gratefully acknowledges financial support from the
New Zealand CoRE Fund via the MacDiarmid Institute and
hospitality at the Cavendish Laboratory, University
of Cambridge during initial stages of this work. J.B.,
M.G.\ and U.Z.\ were supported by the Marsden Fund of New
Zealand (contract no.\ MPF-MAU2505) from government
funding administered by the Royal Society Te Ap\=arangi.
The authors thank Saxon Tobeck-Shine for assistance with
numerical calculations, provided as part of a short-term
research project at Victoria University of Wellington.

\end{acknowledgments}

\appendix

\begin{widetext}

\section{Two conventions for defining the two-body
reduced density matrix}
\label{app:2bRDMconv}

The ultracold-atom and condensed-matter communities
have both used the concept of two-body reduced density
matrices extensively, but they adhere to slightly
different conventions. Here we briefly elucidate the
differences so that relevant results can be more easily
juxtaposed.

The starting point of Yang~\cite{Yang1962} is the
$N$-particle density operator $\hat{\varrho}_N$,
whose zero-temperature form can be expressed in terms of
the $N$-particle ground state $\ket{\Psi_0}$ as
\begin{align}
\hat{\varrho}_N &= \ket{\Psi_0}
\bra{\Psi_0} \nonumber \\[0.2cm] &\equiv \int dx_1\,
dx'_1\, dx_2\, dx'_2\dots dx_N\, dx'_N\,\, \ket{x_1, x_2,
\dots, x_N} \, \varrho_N(x_1, x_2,\dots, x_N; x'_1,
x'_2, \dots, x'_N)\, \bra{x'_1, x'_2, \dots, x'_N} \,\, ,
\end{align}
with matrix elements in real-space representation given
by
\begin{align}
\varrho_N(x_1, x_2, \dots, x_N; x'_1, x'_2, \dots,
x'_N) &= \braket{x_1, x_2, \dots, z_N}{\Psi_0}
\braket{\Psi_0}{x'_1, x'_2, \dots, x'_N}\nonumber
\\[0.2cm] &\hspace{4cm} \equiv \Psi_0(x_1, x_2,\dots,
x_N)\, \big[\Psi_0 (x'_1, x'_2,\dots, x'_N)\big]^* \,\, .
\end{align}
The two-body reduced density matrix is then found as
\begin{subequations}
\begin{align}\label{eq:2bRDM1stMing}
\varrho_2(x_1, x_2; x'_1, x'_2) &= N(N-1) \int dz_3
\dots dz_N\,\, \Psi_0(x_1, x_2, z_3, \dots, z_N)\,
\big[\Psi_0(x'_1, x'_2, z_3, \dots, z_N)\big]^*
\,\, , \\[0.2cm] &= \int dz_3\dots dz_N\,\,
\bra{\Psi_0} c_{x'_1}^\dagger \, c_{x'_2}^\dagger
\ket{z_3, \dots, z_N} \bra{z_3, \dots, z_N} c_{x_2}\,
c_{x_1} \ket{\Psi_0} \,\, , \nonumber \\[0.2cm]
&= \expval{c_{x'_1}^\dagger\, c_{x'_2}^\dagger\, c_{x_2}
\, c_{x_1}}{\Psi_0} \equiv \mathrm{Tr}\left( c_{x_1}\,
c_{x_2}\,\hat{\varrho}_N\, c_{x'_2}^\dagger\,
c_{x'_1}^\dagger\right) \,\, . \label{eq:2bRDM2ndYang}
\end{align}
\end{subequations}
The first-quantized form (\ref{eq:2bRDM1stMing}) is given
explicitly, e.g., in Refs.~\cite{Girardeau2006,
Minguzzi2006}, whereas the definition
(\ref{eq:2bRDM2ndYang}) based on second quantization is
the starting point of Yang~\cite{Yang1962}. The spectral
decomposition of the two-body reduced density matrix
reads~\cite{Yang1962,Koscik2023}
\begin{align}\label{eq:YangSpect}
\varrho_2(x_1, x_2; x'_1, x'_2) = \sum_\alpha
n_\alpha \, \chi_\alpha(x_1, x_2)\, \big[
\chi_\alpha(x'_1, x'_2)\big]^* \,\, ,
\end{align}
in terms of the eigenfunctions $\chi_\alpha(x_1,
x_2)$ that satisfy the eigenvalue equation
\begin{align}\label{eq:YangEigen}
\int dx'_1\, dx'_2 \,\,\,\, \varrho_2(x_1, x_2;
x'_1, x'_2)\,\, \chi_\alpha(x'_1, x'_2) &=
n_\alpha\,\, \chi_\alpha(x_1, x_2) \,\, .
\end{align}

In contrast to Yang's notation,
Leggett~\cite{Leggett2006} defines the zero-temperature
two-body reduced density matrix as
\begin{align}\label{eq:Leg2bRDM}
\rho_2(x_1, x_2; x'_1, x'_2) =
\expval{c_{x_1}^\dagger\, c_{x_2}^\dagger\, c_{x'_2}\,
c_{x'_1}}{\Psi_0} \equiv \left[\varrho_2(x_1,
x_2; x'_1, x'_2)\right]^* \,\, .
\end{align}
See Eq.~(2.4.2) in Ref.~\cite{Leggett2006}, where both
the first-quantized and second-quantized forms are
presented. Interestingly, Leggett gives the spectral
decomposition of $\rho_2(x_1, x_2; x'_1, x'_2)$ in terms
of the eigenfunctions $\chi_\alpha(x_1, x_2)$ of
$\varrho_2(x_1, x_2; x'_1, x'_2)$
[Eq.~(\ref{eq:YangEigen})], i.e., he writes
\begin{align}\label{eq:LeggetSpect}
\rho_2(x_1, x_2; x'_1, x'_2) = \sum_\alpha
n_\alpha\, \big[\chi_\alpha(x_1, x_2)\big]^*\,
\chi_\alpha(x'_1, x'_2) \,\, .
\end{align}
See Eq.~(2.4.8) in Ref.~\cite{Leggett2006}. The
eigenvalue equation relating Leggett's form of the
two-body reduced density matrix to the eigenfunctions
utilized in its spectral decomposition
(\ref{eq:LeggetSpect}) is
\begin{align}\label{eq:LeggettEigen}
\int dx_1\, dx_2 \,\,\,\, \rho_2(x_1, x_2; x'_1,
x'_2)\,\, \chi_\alpha(x_1, x_2) &= n_\alpha
\,\, \chi_\alpha(x'_1, x'_2) \,\, .
\end{align}
Leggett's conventions is widely followed by the
condensed-matter community~\cite{Rampp2022}.

Our definition (\ref{eq:T2bCMgen}) of the T2bCM follows
the one introduced in Ref.~\cite{Thompson2024}, which was
inspired by Leggett's convention. In particular, the
$t=0$ limit of the T2bCM recovers the Leggett form of the
two-body reduced density matrix [Eqs.~(\ref{eq:2bRDMgen})
and (\ref{eq:Leg2bRDM})]. In the present work, we also
adopt the form (\ref{eq:T2bCMdiag}) for the T2bCM's
spectral decomposition that recovers
(\ref{eq:LeggetSpect}) in the $t=0$ limit. Note that this
is at variance with the form of the spectral
decomposition employed in Ref.~\cite{Thompson2024}. As a
result, the eigenfunctions of the T2bCM obtained here are
related by complex conjugation to those defined in
Ref.~\cite{Thompson2024}.

\section{Bose-Fermi mapping and calculation of the
FTG-gas T2bCM}\label{app:FBmappCalc}

The FTG-gas Hamiltonian $H$ contains an infinitely strong
attractive short-range two-body interaction. Via the
Bose-Fermi mapping~\cite{Girardeau2004,Cheon1999}, the
$N$-fermion energy-eigenstate wave functions have the
general form
\begin{align}\label{eq:StatesMap}
\Psi_E(z_1,\,\cdots , z_N) \equiv \braket{z_1,\,
\cdots , z_N}{\Psi_E} &= A(z_1,\,\cdots , z_N)
\,\, \Phi_E(z_1,\,\cdots , z_N) \,\, ,
\end{align}
where $\Phi_E(z_1,\,\cdots , z_N)$ denotes the
energy eigenstate of $N$ noninteracting bosons with the
same eigenvalue $E$, and we introduced the
antisymmetrization function~\cite{Girardeau1965,
Girardeau2004}
\begin{align}\label{eq:antiSymmFunct}
A(z_1,\,\cdots , z_N) &= \prod_{1\le j < k\le N}
\sgn(z_j - z_k) \,\,  .
\end{align}
For the $N$-fermion ground-state wave function,
Eq.~(\ref{eq:StatesMap}) specializes to
\begin{align}
\Psi_0(z_1,\,\cdots , z_N)= \prod_{j = 1}^N \Bigg(
\phi(z_j)\prod_{k = j + 1}^N \sgn(z_j - z_k)\Bigg)\,\, ,
\end{align}
where $\phi(z)$ is the lowest-energy single-particle wave
function that depends on specifics of the external
potential.

Like previous studies of conventional superfluidity in
the  FTG gas~\cite{Girardeau2006,Minguzzi2006}, our
calculations employ first-quantisation methods. For easy
reference, we summarize a few relations involving
many-fermion states,
\begin{subequations}\label{eq:PosRels}
\begin{align}
\ket{z_1,\,\cdots , z_N} &=\frac{1}{\sqrt{N!}}\,\,
c^\dag_{z_1}\cdots\, c^\dag_{z_N} \ket{0} \,\,  , \\
\label{eq:PosRel(b)} 
c^\dag_x \ket{z_3,\,\cdots , z_N} &= \sqrt{N-1}\,
\ket{x, z_3,\,\cdots , z_N} \,\,  , \\
\label{eq:PosRel(c)}
\braket{z_2,\,\cdots , z_N}{c_x | \Psi} &=
\sqrt{N}\, \,\Psi(x,z_2,\,\cdots , z_N)\,\,  .
\end{align}
\end{subequations}
In addition, it will be useful to note the following
properties of the antisymmetrisation function
(\ref{eq:antiSymmFunct}) appearing in the Bose-Fermi
mapping (\ref{eq:StatesMap}),
\begin{subequations}
\begin{align}\label{eq:Aprop(a)}
A(x,z_2,\,\cdots , z_N) &= \Bigg( \prod_{j=2}^N
\sgn(x - z_j)\Bigg) A(z_2,\,\cdots , z_N) \,\,  ,
\\ \label{eq:Aprop(b)}
A(x,z_2,\,\cdots , z_N)\, A(y,z_2,\,\cdots ,
z_N) &= \prod_{j=2}^N \sgn(x - z_j) \,\, \sgn(y - z_j)
\,\, .
\end{align}
\end{subequations}

The general form (\ref{eq:T2bCMgen}) of the T2bCM
involves a $t$-dependent part that mirrors ordinary
quantum time evolution. This analogy is only formal, as
the parameter $t$ embodies the time interval at which
pairing correlations are probed. Nevertheless, it turns
out that quantities of interest can be calculated using
propagator methods.

For the following, the Feynman propagator of the
$(N-1)$-particle FTG gas is relevant. It is given by
\begin{align}
G_\mathrm{FTG}(z_2,\,\cdots , z_N; z'_2,\,\cdots ,
z'_N; t) &\equiv \Bigg\langle z_2,\,\cdots , z_N\Bigg|
\exp(-i\,\frac{t}{\hbar}\, H)\Bigg|z'_2,\,\cdots , z'_N
\Bigg\rangle\nonumber\\[0.2cm] \label{eq:G(b)}
&= \sum_E \exp(-i\,\frac{t}{\hbar}\, E)\, \Psi_E
(z_2,\,\cdots , z_N)\, \big[ \Psi_E(z'_2,\,
\cdots , z'_N)\big]^* \,\, . 
\end{align}
Using the Bose-Fermi-mapping relation
Eq.~(\ref{eq:StatesMap}), we find
\begin{align}\label{eq:Gmap}
G_\mathrm{FTG}(z_2,\,\cdots , z_N; z'_2,\,\cdots ,
z'_N; t) = A(z_2,\,\cdots , z_N)\, A(z'_2,
\,\cdots , z'_N)\,\, G_\mathrm{B}(z_2,\,\cdots ,
z_N; z'_2,\, \cdots , z'_N; t)\,\, ,
\end{align}
where
\begin{align}
G_\mathrm{B}(z_2,\cdots,z_N;z'_2,\cdots,z'_N;t) &=
\sum_E \exp(-i\,\frac{t}{\hbar}\, E)\, \Phi_E
(z_2,\, \cdots , z_N)\, \big[ \Phi_E(z'_2,\,
\cdots , z'_N)\big]^*
\end{align}
is the propagator for the $(N-1)$-particle ideal Bose
gas. Due to the absence of interactions between bosons,
the  $(N-1)$-boson propagator can be written
as~\cite{Brosens1997}
\begin{align}\label{eq:G(B)fac}
G_\mathrm{B}(z_2,\, \cdots , z_N; z'_2,\,\cdots ,
z'_N; t) &= \frac{1}{(N-1)!} \sum_{(i_2 \dots  i_N)}
G(z_2; z'_{i_2}; t)\,\cdots\, G(z_N; z'_{i_N}; t)\,\,  .
\end{align}
Here $(i_2\dots i_N)$ denotes a permutation of the
$(N-1)$ particle indices $(2\dots N)$, and $G(z; z'; t)$
is the noninteracting single-particle propagator.

Our goal is to transform Eq.~(\ref{eq:T2bCMgen}) into a
purely first-quantised form, where the Bose-Fermi mapping
is useful. This proceeds via rewriting the T2bCM as
\begin{align}\label{eq:RhoIntermed}
\rho_2(x,y; x',y';t) &= \int dz_3 \cdots dz_N\,
\braket{\Psi_0}{c^\dag_x\,\exp(-i\,\frac{t}{\hbar}
\, H) \, c^\dag_y \Bigg| z_3\cdots z_N} \braket{z_3\cdots
z_N}{c_{y'}\,\exp(i\,\frac{t}{\hbar}\, H)\,c_{x'} \Bigg|
\Psi_0} \,\, . 
\end{align}
We now consider
\begin{subequations}
\begin{align}\label{eq:propStep1}
&\braket{z_3\cdots z_N}{c_{y'}\,\exp(i\,\frac{t}{\hbar}\,
H)\, c_{x'} \Bigg| \Psi_0} \nonumber\\[0.1cm]
&= \int dy_2\cdots dy_N \braket{z_3,\,\cdots ,
z_N}{c_{y'}\,\exp(i\,\frac{t}{\hbar}\, H) \Bigg| y_2,\,
\cdots , y_N}\braket{y_2,\,\cdots , y_N}{c_{x'} | \Psi_0}
\,\, , \\[0.1cm] \label{eq:propStep2} &= \sqrt{N(N-1)}
\int dy_2\cdots dy_N\,\, G_\mathrm{FTG}(y', z_3,\,
\cdots , z_N; y_2,\,\cdots , y_N; -t) \,\,\Psi_0(x',
y_2,\,\cdots , y_N) \,\, ,\\[0.1cm] \label{eq:propStep3}
&= \sqrt{N(N-1)}\,\,\phi(x')\, A(y',z_3,\,\cdots ,
z_N)\, \int dy_2\cdots dy_N\,\,G_\mathrm{B}(y',
z_3,\,\cdots , z_N; y_2,\,\cdots , y_N; -t) \nonumber\\
& \hspace{8.5cm}\times \sgn(x'-y_2)\,\phi(y_2)\,\,\cdots
\,\,\sgn(x'-y_N)\,\phi(y_N) \,\, , \\[0.1cm]
\label{eq:propStep4}
&= \sqrt{N(N-1)}\,\,\phi(x')\, A(y', z_3,\,\cdots
, z_N)\, \frac{1}{(N-1)!}\sum_{(i_2\cdots i_N)}\,\, \int
dy_2\cdots dy_N \nonumber \\ &\hspace{2cm} G(y'; y_{i_2};
-t)\, G(z_3; y_{i_3}; -t)\,\cdots\, G(z_N; y_{i_N}; -t)
\,\, \sgn(x' - y_2)\, \phi(y_2)\cdots\, \,\sgn(x' - y_N)
\, \phi(y_N) \,\, , \\[0.1cm] \label{eq:propStep5}
&= \sqrt{N(N-1)}\,\,\phi(x')\, A(y', z_3,\,\cdots
, z_N)\, \frac{1}{(N-1)!}\sum_{(i_2\cdots i_N)}\,\, \int
dy_{i_2}\,\, G(y'; y_{i_2}; -t)\, \sgn(x' - y_{i_2})\,
\phi(y_{i_2}) \nonumber  \\ & \hspace{8cm} \times
\prod_{l=3}^N \left[ \int dy_{i_l}\,\, G(z_l; y_{i_l};
-t)\,\sgn(x' - y_{i_l})\,\phi(y_{i_l})\right]\,\, ,
\\[0.1cm] \label{eq:PsiFres}
&= \sqrt{N(N-1)}\,\,\phi(x')\,\varphi_{x'}(y';t)\,
A(y',z_3,\,\cdots , z_N)\, \varphi_{x'}(z_3;t)\,
\cdots\,\varphi_{x'}(z_N;t)\,\, .
\end{align}
\end{subequations}
In (\ref{eq:PsiFres}), we introduced the abbreviation
\begin{align}\label{eq:varPhiDef}
\varphi_x(y; t) = \int dz\,\, G(y; z; -t)\,\, \sgn(x - z)
\,\,\phi(z) \,\, ,
\end{align}
which has the structure of a time-evolved single-particle
wave function that depends parametrically on the fixed
coordinate $x$. To obtain the final form
(\ref{eq:PsiFres}), we made use of
Eqs.~(\ref{eq:PosRel(b)}) and (\ref{eq:PosRel(c)}) [to
get \eqref{eq:propStep2}], the Bose-Fermi mapping of the
FTG propagator (\ref{eq:Gmap}) and relation
(\ref{eq:Aprop(a)}) [to get (\ref{eq:propStep3})], as
well as the form of the noninteracting-boson propagator
(\ref{eq:G(B)fac}) [getting \eqref{eq:propStep4}].
Rewriting (\ref{eq:propStep4}) into (\ref{eq:propStep5})
is facilitated by the ability to rearrange the product of
$\sgn(x'-y_k)$ and $\phi(y_k)$ functions so that it
corresponds to any specific permutation of coordinates
$y_{i_l}$ in the product of single-particle propagators,
leading to the factorization of the multidimensional
integral over coordinates $y_2\dots y_N$. As the same
result is contributed by each term in the sum over
permutations, the sum just cancels the $1/(N-1)!$
prefactor. The form of (\ref{eq:PsiFres}) accords with
the basic intuition for how the BEC state $\ket{\Psi_0}$
should evolve under the noninteracting-boson Hamiltonian.

Using the result from Eq.~(\ref{eq:PsiFres}) in
(\ref{eq:RhoIntermed}) yields
\begin{subequations}
\begin{align}\label{eq:T2bCMstep1}
\rho_2(x, y; x', y'; t) &= N(N-1)\,\phi^*(x)\,
\varphi^*_x(y;t)\,\phi(x')\,\varphi_{x'}(y';t)\nonumber\\
&\hspace{-1cm}\times \int dz_3 \cdots dz_N \,\, A
(y, z_3,\,\cdots , z_N)\, A(y', z_3,\,\cdots , z_N)
\, \varphi^*_x(z_3; t)\,\varphi_{x'}(z_3; t)\,\cdots\,
\varphi^*_x(z_N; t)\,\varphi_{x'}(z_N; t) \,\, ,\\[0.1cm]
\label{eq:T2bCMstep2}
&= N(N-1)\,\phi^*(x)\, \varphi^*_x(y; t)\,\phi(x')\,
\varphi_{x'}(y'; t)\prod_{l=3}^N \left( \int dz_l\,\,
\sgn(y - z_l)\,\sgn(y' - z_l)\, \varphi^*_x(z_l; t)\,
\varphi_{x'}(z_l; t)\right)\,\, .
\end{align}
\end{subequations}
Relation (\ref{eq:Aprop(b)}) was used to obtain the final
form (\ref{eq:T2bCMstep2}). The latter's equivalence to
(\ref{eq:T2bCM1st}) becomes apparent by noting $\varphi_x
(y; t)=e^{i E_0 t/\hbar}\,\psi_x(y; t)$ [compare
Eqs.~(\ref{eq:PsiDef}) and (\ref{eq:varPhiDef})] and the
definition of $P(x, y; x', y'; t)$ from
Eq.~(\ref{eq:PfacDef}).

\section{Dressed single-fermion wave function for
specific external potentials}
\label{app:dressedWF}

\subsection{Homogeneous FTG gas}

Following Ref.~\cite{Girardeau2006}, we assume a finite
system size $L$ and periodic boundary conditions so that
the single-particle eigenstates are $\phi_k(z) = \exp(i k
z)/\sqrt{L}$, with discrete wave numbers $k = n\, 2\pi/L$
and corresponding energies $E_k = \hbar^2 k^2/(2m)$. We
assume $L$ to be large. The homogeneous system's
single-particle propagator then takes the form
\begin{align}\label{eq:freeG}
G(y; z; t) &= \frac{1}{L} \sum_k \,\exp\left[ -i\,
\frac{\hbar\, t}{2m}\, k^2 + i\, (y - z)\, k \right] =
\frac{1}{2\pi}\sum_k\, \Delta k\,\, \exp\left[ -i\,
\frac{\hbar\, t}{2m}\, k^2 + i\, (y - z)\, k\right]\,\, ,
\nonumber \\[0.1cm]
&\approx \frac{1}{2\pi} \int_{-\infty}^\infty dk \,\,
\exp\left[-i\, \frac{\hbar\, t}{2m}\, k^2 + i (y - z) k
\right] \,\, , \nonumber \\[0.1cm]
&= \sqrt{\frac{m}{2\pi i \hbar t}}\,\exp\left[i\,
\frac{m}{2\hbar t}\,(y-z)^2\right]\,\, . 
\end{align}
Here we approximated $\sum_k$ as an integral because it
has the form of a Riemann sum, with $\Delta k = 2\pi/L
\to 0$ when $L \to \infty$. The result (\ref{eq:freeG})
for the propagator agrees with that of a free particle in
an infinite system given, e.g., in Eq.~(2.6.16) from
Ref.~\cite{Sakurai2011}.

We now use $\phi(z) = 1/\sqrt{L}$ (the eigenstate
corresponding to the lowest-energy eigenvalue $E_0=0$)
and the form (\ref{eq:freeG}) for the propagator in
Eq.~(\ref{eq:PsiDef}). This yields
\begin{subequations}
\begin{align}
\psi_x(y; t) &= \frac{1}{\sqrt{L}}\,\sqrt{\frac{m}{2\pi i
\hbar(-t)}}\,\int_{-\frac{L}{2}}^{\frac{L}{2}} dz\, \exp
\left[i\, \frac{m}{2\hbar (-t)}\,(y-z)^2\right]\,\,
\sgn(x - z) \,\, , \\[0.1cm]
&= \frac{1}{\sqrt{L}}\,\frac{1}{\sqrt{2\pi(-i)\omega_L
t}}\,\,\int_{-\frac{1}{2}}^{\frac{1}{2}} d\left(
\frac{z}{L}\right)\,\, \exp\left[\frac{-i}{2\omega_L t}
\,\left(\frac{y}{L} - \frac{z}{L}\right)^2\right]\,\,
\sgn\left(\frac{x}{L} - \frac{z}{L}\right) \,\, ,
\\[0.1cm]
&= \frac{1}{\sqrt{L}}\,\frac{1}{\sqrt{2\pi(-i)\omega_L
t}}\,\left[ \int_{-\frac{1}{2}-
\frac{y}{L}}^{\frac{x-y}{L}} d\left(\frac{z}{L}\right)
-\int_{\frac{x-y}{L}}^{\frac{1}{2}-\frac{y}{L}} d\left(
\frac{z}{L}\right) \right]\, \exp\left[ \frac{-i}{2
\omega_L t}\,\left(\frac{z}{L}\right)^2\right]\,\, ,
\\[0.1cm] \label{eq:PsiFreeStep1}
&= \frac{1}{\sqrt{L}}\,\,\mathrm{erf}\left(
\sqrt{\frac{i}{2\omega_L t}}\, \frac{x-y}{L} \right) +
\frac{1}{2\sqrt{L}}\left\{ \mathrm{erf}\left[
\sqrt{\frac{i}{2\omega_L t}}\left(\frac{1}{2}+
\frac{y}{L}\right)\right] - \mathrm{erf}\left[
\sqrt{\frac{i}{2\omega_L t}}\left(\frac{1}{2}
-\frac{y}{L}\right)\right]\right\}\,\, .
\end{align}
\end{subequations}
For $\omega_L |t|\ll 1$ and $|y|< L/2$, the boundary
terms in (\ref{eq:PsiFreeStep1}) almost perfectly cancel
and (\ref{eq:FreeGasPsi}) is obtained.

\subsection{FTG gas in a harmonic-oscillator potential}

The closed-form expression for the harmonic-oscillator
propagator is~\cite{Sakurai2011,Holstein1998,
Thornber1998,Chaos1999}
\begin{align}\label{eq:oscProp}
G(y; z; t) = \frac{1}{l_\omega}\,\frac{1}{\sqrt{2\pi i
\hbar\sin(\omega t)}}\,\,\exp\left[\frac{i}{2}\,
\frac{\left( y^2 + z^2\right)\cos(\omega t) -2\, y\,
z}{l_\omega^2\, \sin(\omega t)} \right]\,\, .
\end{align}
Using the oscillator ground-state wave function $\phi(z)
=\exp[-z^2/(2\,l_\omega^2)]/\sqrt{\sqrt{\pi}\,l_\omega}$,
its energy eigenvalue $E_0=\hbar\omega/2$, and the
propagator from Eq.~(\ref{eq:oscProp}) in the definition
(\ref{eq:PsiDef}) for the dressed single-fermion wave
function, we find
\begin{align*}
\psi_x(y; t) &= e^{-i\frac{\omega t}{2}}\,\,
\sqrt{\frac{i}{2\pi^{3/2}l_\omega^3 \sin(\omega t)}}\,\,
\int_{-\infty}^\infty dz\,\exp\left[-
\frac{1}{2 l_\omega^2}\left( z^2 + i\,\frac{y^2 +
z^2}{\tan(\omega t)} - 2 i\,\frac{y\, z}{\sin(\omega t)}
\right)\right]\, \sgn(x - z)\,\, ,\\[0.1cm]
&= e^{-i\frac{\omega t}{2}}\,\,\sqrt{\frac{i}{2\pi^{3/2}
l_\omega^3 \sin(\omega t)}}\,\,\exp\left(-i\,\frac{y^2}{2
l_\omega^2\tan(\omega t)}\right) \nonumber \\[0.1cm]
& \hspace{5cm} \times \int_{-\infty}^\infty dz\,\exp
\left[-\frac{1}{2l_\omega^2}\left( [1 + i\cot(\omega t)]
z^2 - 2 z\, \frac{i y}{\sin(\omega t)} \right) \right]\,
\sgn(x - z) \,\, .
\end{align*}
Completing the square and noting $1 + i\,\cot(\omega t) =
i\, e^{-i\omega t}/\sin(\omega t)$ gives
\begin{align}
\psi_x(y; t) &= e^{-i\frac{\omega t}{2}}\,\,
\sqrt{\frac{i}{2\pi^{3/2}l_\omega^3 \sin(\omega t)}}\,\,
\exp\left( -\frac{y^2}{2\,l_\omega^2}\right)\,
\int_{-\infty}^{\infty} dz\, \exp\left[ -\frac{i\,e^{-i
\omega t}}{2\,l_\omega^2\sin(\omega t)} \left( z - e^{i\,
\omega t}\, y \right)^2 \right] \sgn(x - z) \,\, ,
\nonumber \\[0.1cm]
&= e^{-i\frac{\omega t}{2}}\,\,\sqrt{\frac{i}{2\pi
l_\omega^2 \sin(\omega t)}}\,\,\,\phi(y)\,
\int_{-\infty}^\infty dw\,\, \exp\left[ -\frac{i\,
e^{-i\,\omega t}}{2 l_\omega^2\sin(\omega t)} \, \left(
x - e^{i\,\omega t}\, y - w\right)^2\right]\,\sgn(w)
\,\, , \nonumber \\[0.1cm]  \label{eq:HarmPsiStep}
&= \phi(y)\,\,\erf\left[ \sqrt{\frac{i}{2 l_\omega^2 \sin
(\omega t)}}\left( e^{-i\frac{\omega t}{2}}\, x
- e^{i\frac{\omega t}{2}}\, y\right) \right] \,\, .
\end{align}
In the last step to obtain (\ref{eq:HarmPsiStep}), we
used the identity (see 3.546~1 in
Ref.~\cite{Gradshteyn1994})
\begin{align*}
\int_{-\infty}^{\infty} dw\,\,\sgn(w)\, e^{-C\,(w_0
- w)^2} = 2 e^{-C w_0^2} \int_0^\infty dw\,\, e^{-C\,
w^2} \sinh\left( 2 C w_0\, w \right) =
\sqrt{\frac{\pi}{C}}\, \erf\left(\sqrt{C}\,w_0\right)
\,\, .
\end{align*}

\begin{figure*}[t]
\includegraphics[height=3.7cm]{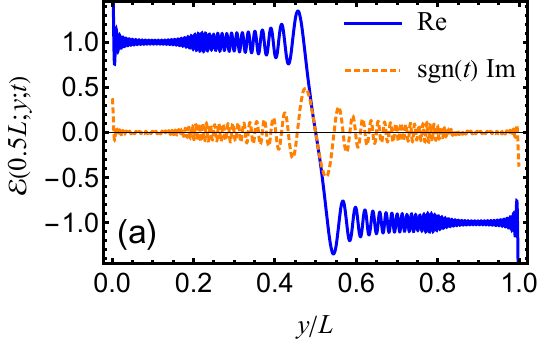}
\hspace{1cm}
\includegraphics[height=3.7cm]{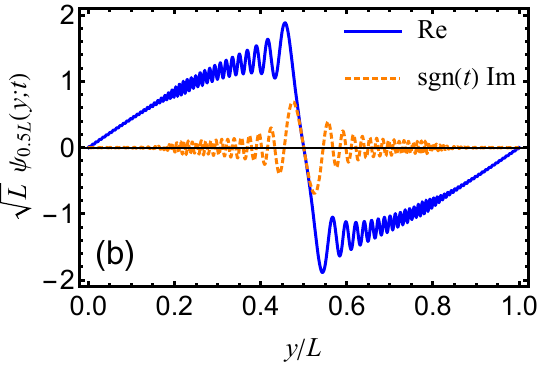}
\caption[]{\label{fig:SqWellPsi}%
(a)~Propagated signum function $\mathcal{E}(x; y; t)$
[Eq.~(\ref{eq:SqWellErf})] for $x=0.5 L$ and $\omega_0 t 
= 0.002$. (b)~Dressed single-fermion wave function
$\psi_x(y;t)$ of the square-well-confined FTG gas
[Eq.~(\ref{eq:SqWellPsi})] for $x=0.5 L$ and $\omega_0 t 
= 0.002$.}
\end{figure*}

\subsection{FTG gas in a square-well potential}

The propagator for the infinite square-well potential is
only known in terms of a series~\cite{Gori2001,
Fulling2003};
\begin{equation}
G(y;z;t) = \frac{2}{L}\,\sum_{n=1}^\infty \sin\left(n\pi
\, \frac{y}{L}\right)\,\sin\left(n\pi\,\frac{z}{L}
\right)\,\, e^{-i n^2 \omega_0 t} \quad .
\end{equation}
Here $L$ is the well width, and $\omega_0\equiv E_0/
\hbar=\pi^2\hbar/(2 m L^2)$ denotes the frequency scale
for the well ground state whose wave function is
$\phi(z)=\sqrt{2/L}\,\sin(\pi z/L)$.

Calculation of the dressed single-fermion wave function
as per its definition (\ref{eq:PsiDef}) for square-well
confinement proceeds over the following steps:
\begin{align*}
\psi_x(y;t) &= \sqrt{\frac{2}{L}}\,\, 2\,
\sum_{n=1}^\infty e^{i (n^2 - 1)\omega_0 t} \,\, \sin
\left(n\pi\,\frac{y}{L}\right)\, \int_0^L \frac{dz}{L}
\,\, \sin\left(n\pi\,\frac{z}{L}\right)\,\,\mathrm{sgn}
(x-z)\,\sin\left(\pi\,\frac{z}{L}\right)\,\, , \\[0.1cm]
&= \sqrt{\frac{2}{L}}\,\, 2\,\sum_{n=1}^\infty e^{i (n^2
- 1)\omega_0 t} \,\, \sin\left(n\pi\,\frac{y}{L}\right)
\left\{ \int_0^x - \int_x^L\right\}\frac{dz}{L}\,\,
\sin\left(n\pi\,\frac{z}{L}\right)\,\,\sin\left(\pi\,
\frac{z}{L}\right)\,\, .
\end{align*}
Making use of the definite integrals
\begin{align*}
\int^x \frac{dz}{L}\,\,\sin^2\left(\pi\,\frac{z}{L}
\right) &= \frac{x}{2 L} - \frac{1}{4\pi}\,\sin\left(
2\pi\, \frac{x}{L} \right) \,\, , \\[0.1cm]
\int^x \frac{dz}{L}\,\, \sin\left(n\pi\,\frac{z}{L}
\right)\,\,\sin\left(\pi\,\frac{z}{L}\right) &=
\frac{1}{\pi(n^2-1)}\left[ \sin\left(n \pi\frac{x}{L}
\right) \cos\left(\pi\frac{x}{L}\right)  - n\, \cos
\left(n \pi \frac{x}{L}\right)\sin\left(\pi \frac{x}{L}
\right)\right]\,\, , \\[0.1cm]
&=\frac{1}{2\pi}\left[\frac{\sin\left[(n-1)\pi\frac{x}{L}
\right]}{n-1} - \frac{\sin\left[(n+1)\pi\frac{x}{L}
\right]}{n+1}\right] \,\, ,
\end{align*}
we find
\begin{subequations}
\begin{align}\label{eq:SqWellPsi}
\psi_x(y; t) &= \phi(y)\,\,\mathcal{E}(x;y;t)\,\, ,
\\[0.1cm] \label{eq:SqWellErf}
\mathcal{E}(x; y; t) &= \frac{2 x}{L}-\frac{1}{\pi}\,\sin
\left(2\pi\frac{x}{L}\right) - 1 + \frac{2}{\pi}\,
\sum_{n=2}^\infty \, e^{i (n^2 - 1) \omega_0 t}\,\,
\frac{\sin\left(n\pi\frac{y}{L}\right)}{\sin\left(\pi
\frac{y}{L}\right)} \left[ \frac{\sin\left[(n-1)\pi
\frac{x}{L}\right]}{n-1} - \frac{\sin\left[(n+1)\pi
\frac{x}{L}\right]}{n+1}\right]\,\, .
\end{align}
\end{subequations}
The line shape of $\mathcal{E}(x; y; t)$ is reminiscent
of an error function. We illustrate the coordinate
dependences of both $\mathcal{E}(x; y; t)$ and $\psi_x(y;
t)$ in Fig.~\ref{fig:SqWellPsi}.

\section{Macroscopic eigenvalue of the homogeneous FTG
gas at finite \texorpdfstring{$\bm{t}$}{$t$}}
\label{app:ThetaFuncs}

\begin{figure*}
\includegraphics[height=3.8cm]{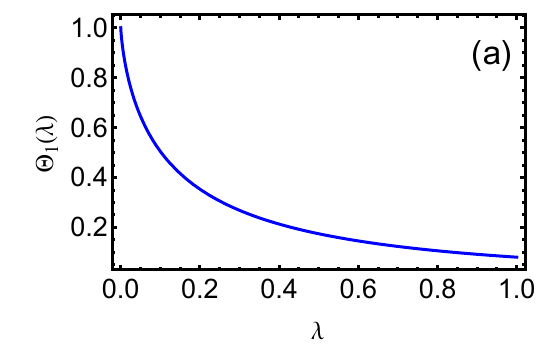}
\hfill
\includegraphics[height=3.8cm]{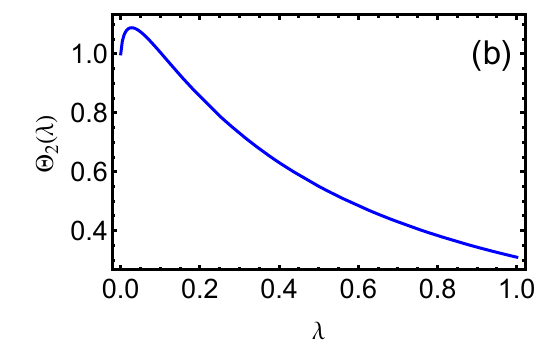}
\hfill
\includegraphics[height=3.8cm]{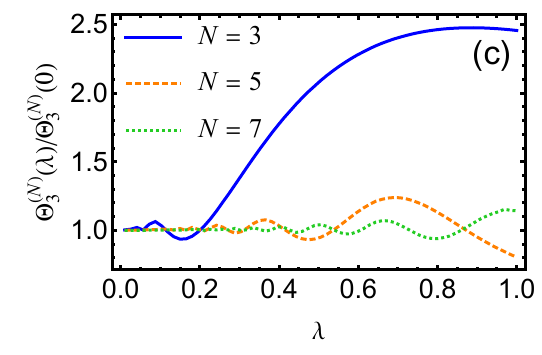}
\caption{\label{fig:FreeGasNormTs}%
Functions governing all $t$ dependence in the
normalization factor (\ref{eq:CNfinal}) and the
macroscopic eigenvalue (\ref{eq:FreeGasN0}) for the
homogeneous FTG gas.} 
\end{figure*}

In terms of universal coordinates $\xi=F(x)$ and $\eta=
F(y)$, the expression for the normalization factor of the
homogeneous-FTG-gas ODLRO eigenfunction becomes
\begin{align}
\left|\mathcal{C}_t\right|^{-2} &= \int_0^1 d\xi\,
\int_0^1 d\eta \,\left|\mathrm{erf}\left(\sqrt{\frac{i}{2
\omega_L t}}\,[\xi-\eta]\right) \right|^2 e^{-4(N-2) |\xi
- \eta|} \,\,e^{-\frac{2}{\sqrt{\pi}}(N-2)|\xi-\eta|\,
\Re\left\{\Gamma\left( -\frac{1}{2},\frac{i}{2\omega_L t}
[\xi-\eta]^2\right)\right\}}\,\, , \\[0.1cm]
\label{eq:CNstep}
&= 2 \int_0^1 d\xi \,\, \int_0^\xi d\eta \,\, \left|
\mathrm{erf}\left(\sqrt{\frac{i}{2\omega_L t}}\,\eta
\right)\right|^2 e^{-4(N-2)\eta}\,\,
e^{-\frac{2}{\sqrt{\pi}}(N-2)\eta\, \Re\left\{\Gamma
\left(-\frac{1}{2},\frac{i}{2\omega_L t}\, \eta^2
\right)\right\}} \,\, , \\[0.2cm] \label{eq:CNfinal}
&= \frac{1}{2(N-2)}\,\,\Theta_1\left([N-2]^2\omega_L t
\right) - \frac{1}{8(N-2)^2} \left\{ \Theta_2\left([N
-2]^2\omega_L t\right) - \Theta_3^{(N)}\left([N-2]^2
\omega_L t \right)\right\}\,\, .
\end{align}
The $t$-dependent renormalisation factors appearing in
(\ref{eq:CNfinal}) are (see also plots in
Fig.~\ref{fig:FreeGasNormTs})
\begin{subequations}
\begin{align}
\Theta_1(\lambda) &= \int_0^\infty d\eta \,
\left|\mathrm{erf} \left(\sqrt{\frac{i}{2\lambda}}\,
\frac{\eta}{4}\right)\right|^2 \exp(-\eta\left[1 +
\frac{1}{2\sqrt{\pi}}\,\Re\left\{\Gamma\left(
-\frac{1}{2},\frac{i}{2\lambda}\,\frac{\eta^2}{16}
\right)\right\}\right])\,\, , \\[0.1cm]
\Theta_2(\lambda) &= \int_0^\infty d\xi\,\,
\int_\xi^\infty d\eta\,\left|\mathrm{erf}\left(
\sqrt{\frac{i}{2\lambda}}\,\frac{\eta}{4}\right)
\right|^2 \exp(-\eta\left[1 + \frac{1}{2\sqrt{\pi}}\,
\Re\left\{\Gamma\left(-\frac{1}{2},\frac{i}{2\lambda}\,
\frac{\eta^2}{16}\right)\right\}\right])\,\, ,\\[0.1cm]
\Theta_3^{(N)}(\lambda) &= \int_{4(N-2)}^\infty d\xi\,
\int_\xi^\infty d\eta\, \left|\mathrm{erf}\left(
\sqrt{\frac{i}{2\lambda}}\,\frac{\eta}{4}\right)
\right|^2 \exp(-\eta\left[1 + \frac{1}{2\sqrt{\pi}}\,
\Re\left\{\Gamma\left( -\frac{1}{2}, \frac{i}{2\lambda}\,
\frac{\eta^2}{16}\right)\right\}\right])\,\, .
\end{align}
\end{subequations}
Using the result (\ref{eq:CNfinal}) in
(\ref{eq:ODLROeval}) yields Eq.~(\ref{eq:FreeGasN0}).

\end{widetext}

\section{Scaling analysis for the odd-frequency-pairing
magnitude}
\label{app:scaling}

\begin{figure*}
\includegraphics[width=0.34\textwidth]{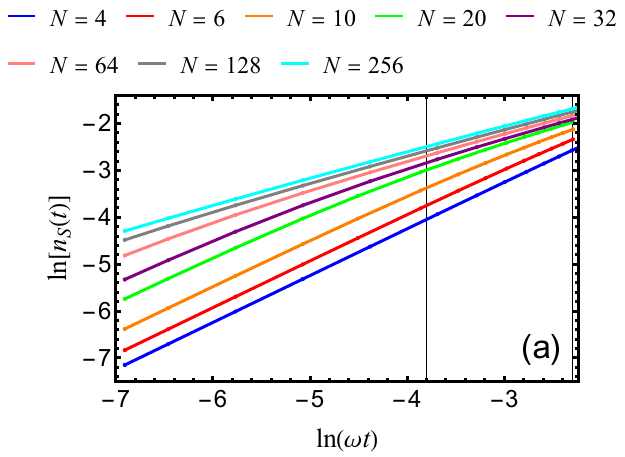}\hfill
\includegraphics[width=0.32\textwidth]{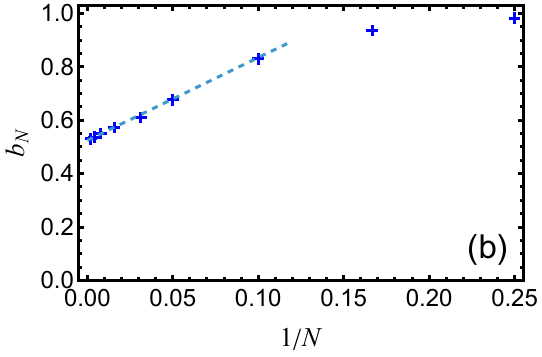}\hfill
\includegraphics[width=0.325\textwidth]{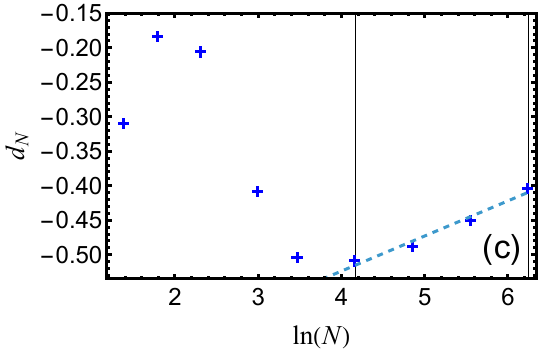}
\caption{\label{fig:scaling}%
Scaling analysis for the odd-frequency pairing magnitude
as per the \textit{Ansatz}~\eqref{eq:nS(t)scale}.
(a)~Log-log plot of $n_S(t)$ as a function of $\omega t$
for fixed values of $N$ indicated in the legend. We
focus on the linear regime delimited by the vertical
gray lines that is representative of the fixed-$t$,
large-$N$ limit relevant for discussing macroscopicity.
(b)~Slopes $b_N$ extracted from linear fits of $\ln(n_S)$
vs.\ $\ln(\omega t)$ in the region delimited by gray
lines in panel~(a), plotted as a function of $1/N$.
Linear extrapolation of $b_N$ for $1/N\to 0$ yields our
quoted value for the exponent $b$. (c)~Intercepts $d_N$
extracted from linear fits of $\ln(n_S)$ vs.\ $\ln(\omega
t)$ in the region delimited by gray lines in panel~(a),
plotted as a function of $\ln N$. An approximately linear
trend emerges for $N\gtrsim 64$, i.e., particle numbers
for which the curves in panel (a) are in regime (ii) over
the entire fitting range. Data points in the range
delimited by the gray vertical lines are fitted to the
linear relationship \eqref{eq:dNscale} to obtain $a$ and
$\ln c$. Dashed lines in panels (b) and (c) are plots of
linear fits to selected data.}
\end{figure*}

The analysis yielding the results quoted in the main
text is based on consideration of the log-log version of
the scaling \textit{Ansatz}~\eqref{eq:nS(t)scale},
\begin{align}\label{eq:nS(t)scaleLog}
\ln(n_S) &= b\,\ln(\omega t)\, +\,\ln(c\, N^a)\,\, .
\end{align}
See Fig.~\ref{fig:scaling}(a) for a plot of this relation
for a subset of data used in the scaling analysis. For
each curve (corresponding to a given $N$), we can
identify two linear regimes separated by the time scale
$t_\mathrm{c}(N)\propto(N-2)^{-2/3}\,\omega^{-1}$. This
time scale emerges from $\Re\{p(x,y;t)\}-\Re\{p(x,y;0)\}
\propto(\omega t)^{3/2}$ together with
Eq.~\eqref{eq:PexpoN} to leading order for $\omega t\to
0$. Regime (i) is defined by $t < t_\mathrm{c}(N)$ and
corresponds to the case where $n_S(t)$ is linear in
$\omega t$ for all $N$ [see also Fig.~\ref{fig:nS(t)}].
Regime (ii) is characterised by $N$-dependent
nonlinearity in $n_S(t)$ and is defined by $t_\mathrm{c}
(N) < t < 0.1\,\omega^{-1}$. Linearity of $n_S(t)$ is
actually expected to arise only for regime (i) because
the quantity $|S\psi_x(y;t)|^2$ entering
Eq.~(\ref{eq:nS(t)def}) via Eq.~(\ref{eq:phiNgen}) is
$\propto\omega t$ and, for times $t < t_\mathrm{c}(N)$,
solely determines the $t$ dependence of $n_S(t)$ due to
$\Re\{p(x,y;t)\}-\Re\{p(x,y;0)\}$ being negligible.
Because $t_\mathrm{c}(N) \to 0$ in the thermodynamic
limit of interest here, we focus on extracting scaling
exponents of $n_S(t)$ in regime (ii) where $t_\mathrm{c}
(N) < t < 0.1\,\omega^{-1}$. For each curve $\ln(n_S)$
vs.\ $\ln(\omega t)$ at fixed $N\in\{4, 6, 10, 20, 32,
64, 128, 256, 512\}$, we fit the portion $-3.80 \le
\ln(\omega t) \le -2.30$ [corresponding to the region
delimited by the two gray vertical lines in
Fig.~\ref{fig:scaling}(a)] to the form
\begin{align}\label{eq:NscaleForm}
\ln(n_S)=b_N\,\ln(\omega t)\, +\, d_N \,\, .
\end{align}
This yields a set of 9 values each for the slope $b_N$ and
intercept $d_N$. Note that $\ln[\omega t_\mathrm{c}(N)]$
grows more negative as $N$ is increased. Larger values of
$N$ therefore ensure that the fitting region in
Fig.~\ref{fig:scaling}(a) covers more of the interval
$\ln[\omega t_\mathrm{c}(N)] < \ln(\omega t) < -2.30$
pertaining to regime (ii) and less of the interpolation
interval $\ln(\omega t) \lesssim \ln[\omega t_\mathrm{c}
(N)]$ that is irrelevant to our fitting analysis. For the
particular choice $-3.80 \le \ln(\omega t) \le -2.30$ for
the fitting region, Fig.~\ref{fig:scaling}(a) suggests
that scaling results will be most meaningful for $N
\gtrsim 64$.

Figure~\ref{fig:scaling}(b) plots the values $b_N$,
obtained for the 9 values of $N$ by the method described
above, as a function of $1/N$. A clear linear trend is
observed for large $N$, i.e., for $1/N\to 0$. We use
linear extrapolation of $b_N$ for $1/N\to 0$ to extract
the value of $b$ quoted in the main text.

The intercepts $d_N$ contain still more information.
Assuming the scaling Ansatz~\eqref{eq:nS(t)scale}, we
expect the relation
\begin{align}\label{eq:dNscale}
d_N = a\,\ln N + \ln c   
\end{align}
to hold. Thus a linear fit of the values $d_N$ as a
function of $\ln N$ should yield $a$ as the slope and
$\ln c$ as the intercept. As seen in
Fig.~\ref{fig:scaling}(c) and expected on the basis of
Fig.~\ref{fig:scaling}(a), only the four data points for
$N\gtrsim 64$ fall approximately on a linear curve. We
fit the $d_N$ values for $N\in\{64, 128, 256, 512\}$ to a
linear relationship as a function of $\ln N$, which
yields the values for $a$ and $c$ given in the main text.


\begin{thebibliography}{86}
\bibitem [{misc:PexpoApprox()}]{misc:PexpoApprox}%
  \BibitemOpen
  \bibinfo {note} {To ensure validity of the approximation $P(x, y; x', y';
  t)\approx e^{- p^*(x, y; t)}\,\, e^{-p(x', y'; t)}$ that underpins
  Eq.~(\ref{eq:PexpoN}), it is essential to adhere to the condition
  (\ref{eq:genODLRO}). In particular, for fixed $x'=y'=0$, we found deviations
  to be minimized for $x=y\ge\ell_\omega$ (i.e., at the trap boundary) and
  bounded by $\omega t/\pi$. Thus, the error inherent in the approximation
  (\ref{eq:PexpoN}) has only a minor effect on the strong boundary-localized
  features visualized, e.g., in Fig.~\ref{fig:ODLROop}.}\BibitemShut {Stop}%
\end{thebibliography}
%
%

\end{document}